# Practitioner forecasts of technological progress in biostasis

Andrew T. McKenzie[1*], Michael Cerullo[2,3], Navid Farahani[3], Jordan S. Sparks[1], Taurus Londoño[4], Aschwin de Wolf[5,6], Suzan Dziennis[5], Borys Wróbel[7,8,9], Alexander German[10], Emil F. Kendziorra[11], João Pedro de Magalhães[12], Wonjin Cho[13], R. Michael Perry[13], Max More[6]

[1] Apex Neuroscience
[2] Carboncopies Foundation
[3] Brain Preservation Foundation
[4] Weill Cornell Medical College
[5] Advanced Neural Biosciences
[6] Biostasis Technologies
[7] European Institute for Brain Research
[8] BioPreservation Institute
[9] Nectome
[10] Department of Molecular Neurology, University Hospital Erlangen
[11] European Biostasis Foundation
[12] Department of Inflammation and Ageing, College of Medicine and Health, University of Birmingham
[13] Alcor Life Extension Foundation

**Abstract**: Biostasis has the potential to extend human lives by offering a bridge to powerful life extension technologies that may be developed in the future. However, key questions in the field remain unresolved, including which biomarkers reliably indicate successful preservation, what technical obstacles pose the greatest barriers, and whether different proposed revival methods are theoretically feasible. To address these gaps, we conducted a collaborative forecasting exercise with 22 practitioners in biostasis, including individuals with expertise in neuroscience, cryobiology, and clinical care. Our results reveal substantial consensus in some areas, for example that synaptic connectivity can serve as a reliable surrogate biomarker for information preservation quality. Practitioners identified three most likely failure modes in contemporary biostasis: inadequate preservation quality even under ideal conditions, geographic barriers preventing timely preservation, and poor procedural execution. Regarding revival strategies, most respondents believe that provably reversible cryopreservation of whole mammalian organisms is most likely decades away, with provably reversible human cryopreservation expected even later, if it is ever achieved. Among revival strategies from contemporary preservation methods, whole brain emulation was considered the most likely to be developed first, though respondents were divided on the metaphysical question of whether it could constitute genuine revival. Molecular nanotechnology was viewed as nearly as likely to be technically feasible, and compatible with both pure cryopreservation and aldehyde-based methods. Taken together, these findings delineate current barriers to high-quality preservation, identify future research priorities, and provide baseline estimates for key areas of uncertainty.

\* Correspondence: amckenzie@apexneuro.org

**Contents**

# Introduction

Biostasis is the preservation of humans for the long-term with the intent of future recovery, if this ever becomes possible. Methods for biostasis include cryopreservation, aldehyde preservation, and their combination. Over the past several years, this field has seen growing research interest (Best, 2008; McKenzie et al., 2024c). Last year, a roadmap was published outlining future directions for biostasis research (McKenzie et al., 2024b). This roadmap discusses potential areas of improvement in biostasis research and practice, including pre-cardiac arrest factors, preservation methods, ways of assessing preservation quality, and possible methods for revival. However, the roadmap did not attempt to appraise the degree of disagreement among practitioners in the field on these topics, nor did it attempt to forecast future technological progress in biostasis.

In this manuscript, we aimed to address these gaps through a forecasting exercise that captures practitioner views on the current state of the field and its future directions. To do so, we first sought opinions from people in the field who are doing or have done direct work in biostasis, at different organizations. We created a questionnaire to capture the views of biostasis *practitioners*, which is our term to describe people with current or previous roles within biostasis research, clinical care, administration, funding/investment, and advocacy. Respondents provided their perspectives on key topics including quality metrics, technical challenges, the comparative potential of different methods, and the feasibility of various revival technologies. In this manuscript, we analyze and discuss these responses. Our goal is to identify areas of consensus and divergence among practitioners, foster collaboration, inspire further debate in areas of disagreement, and clarify the most significant challenges that must be addressed to advance the field of biostasis. We hope that this may be useful for researchers considering entering the field, funding organizations evaluating proposals, and policymakers considering regulatory frameworks.

At the outset, we note some disclaimers. We acknowledge that the practice of biostasis has a high degree of uncertainty, because it relies upon the development of revival technologies that do not currently exist and must be invented in the future. We also acknowledge there is good reason to believe that practitioners in the field will be more optimistic about it than the scientific community at large, due to selection bias. However, many respondents expressed substantial reservations about various aspects of biostasis research. For example, some questioned the validity of surrogate biomarkers as indicators of information preservation, especially in the absence of better validation, such as memory decoding experiments. Others doubted the feasibility of developing provably reversible preservation procedures, especially in the near term. Our goal is to document respondents' honest opinions, and thereby to help identify which questions in this field are the most relevant, important, uncertain, and amenable to testing.

# Methods

We conducted an online questionnaire to gather the opinions of practitioners in biostasis. It was developed in collaboration with co-authors of the previous biostasis roadmap article (McKenzie et al., 2024b). The questionnaire procedures followed the institutional policies of Apex Neuroscience. We gathered respondents by (a) asking the co-authors of the biostasis roadmap article to identify individuals in their networks who might be appropriate and (b) asking the speakers from the Biostasis Week Conference 2025 at Vitalist Bay if they were interested in completing the questionnaire ("Biostasis Week Conference (May 17-18) @ Vitalist Bay," 2025). We also offered respondents the ability to forward the questionnaire to other practitioners in their networks. The questionnaire consisted of multiple-choice questions, numeric probability estimates, and optional open-ended responses covering four main categories: (a) measuring preservation quality, (b) challenges and obstacles in biostasis, (c) preservation methods, and (d) revival technologies. For questions involving probability estimates, respondents were asked to provide their best estimate on a 0-100% scale. Participation was voluntary, and respondents could skip any questions they did not feel qualified to answer or did not wish to answer for any other reason. The questionnaire was conducted in April-June of 2025, and anonymized results were compiled for analysis.

**Respondent background**

*Primary role and length of time engaged in biostasis*

We sent invitations to participate in this project to n = 46 people and received responses from n = 22, for a response rate of 48%. We first asked respondents about their primary role(s) in biostasis. They were allowed to choose multiple roles. The primary roles selected included research (77% of respondents), clinical care (i.e. biostasis procedures; 41%), administration/management (32%), technical/engineering (27%), and funding/investment (23%; **Figure 1a**). Most respondents reported a substantial length of experience in the field, with 73% having been meaningfully engaged with biostasis for more than 5 years, and 59% for 10 or more years (**Figure 1b**).

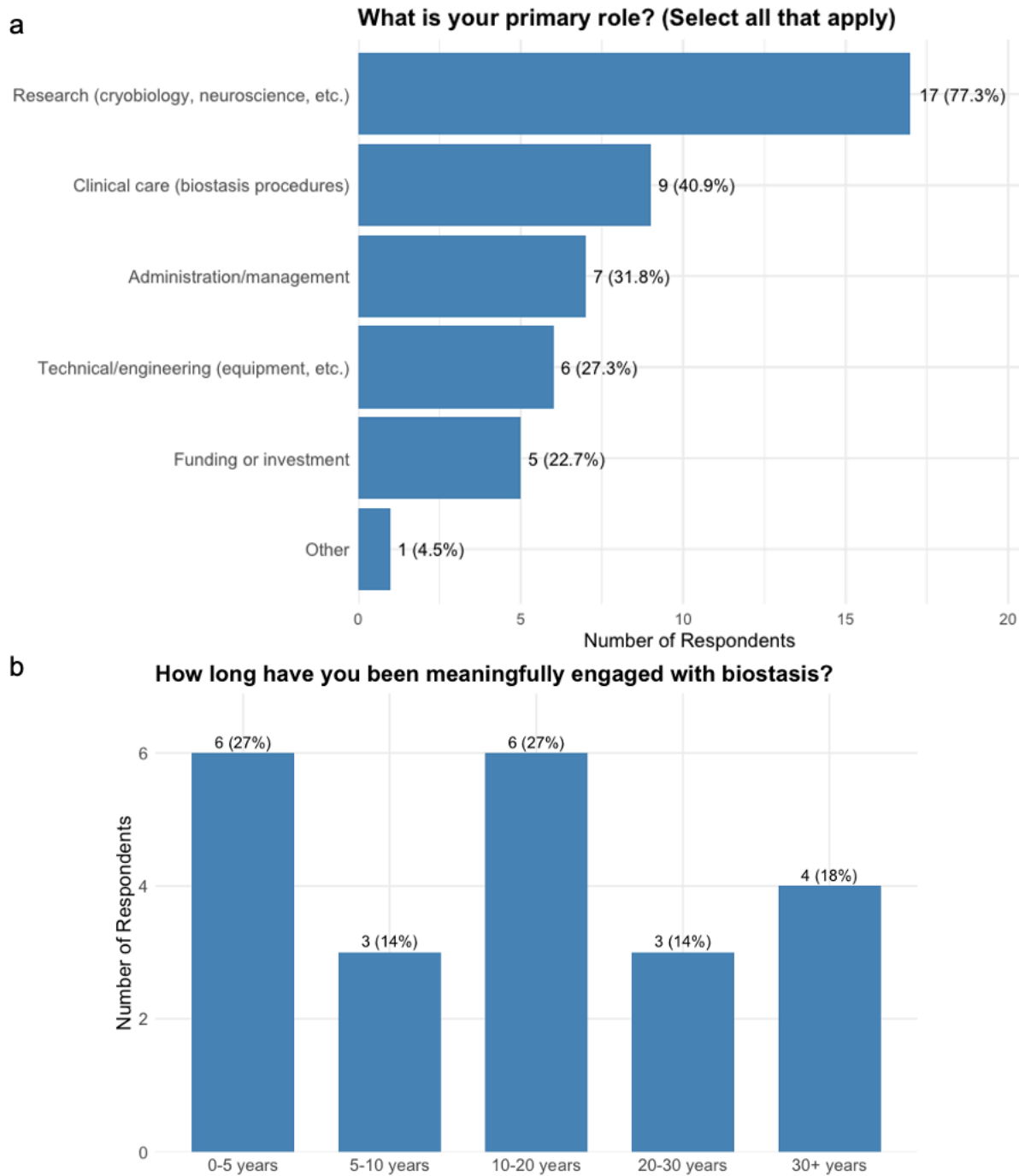

**Figure 1.** Respondent demographics. a: Primary roles of respondents. Respondents were invited to select multiple roles if they each applied. b: Self-reported years of meaningful engagement with biostasis.

*Focus on structural preservation or provably reversible preservation*

We asked respondents the following:

> Evaluating methods for biostasis requires a way to assess their quality. Two distinct approaches have emerged in biostasis research: those focused on preserving surrogate biomarkers that indicate information preservation, and those aiming to develop provably reversible preservation techniques.
>
> The surrogate biomarker approach focuses on the long-term preservation of measurable aspects of the body (especially the brain) believed to encode personal identity, such as long-term memories. The idea here is that while we cannot demonstrate complete reversibility with current technology, we may be able to preserve the information necessary for future revival, if we speculate that technology is likely to improve in the future. One of the major biomarkers that has emerged here is brain structure, especially connectome traceability – i.e., the ability to follow neural connections in 3D on ultrastructural imaging. Another major biomarker that has been proposed is electrophysiology, for example measured on brain slices or isolated brains. Other potential options for biomarkers are cellular viability indicators and molecular integrity.
>
> Provably reversible preservation aims to develop methods where the long-term preservation process can be reversed today, returning the organism or isolated brain (via transplantation) to normal functioning with retention of pre-preservation memories and other aspects of cognition. This approach provides immediate feedback on preservation quality without questions about whether the measurements truly indicate information preservation or whether revival capabilities will ever be developed in the future. While provably reversible preservation has been achieved with small organisms like nematodes, this milestone has not yet been achieved in mammals.
>
> Which of these areas is your work and interest primarily focused on?

Of the respondents, 10 (50%) reported they focus mostly or exclusively on surrogate biomarker preservation, 6 (30%) focus mostly or exclusively on provably reversible preservation, and 4 (20%) focus on both approaches roughly equally.

*Focus on ideal or less-than-ideal conditions*

We asked respondents the following:

> In biostasis, the biological state of the patient at the time of preservation greatly affects the potential for future revival. We can distinguish between two main types of scenarios – ideal conditions and less-than-ideal conditions, as described in (Wowk, 2022).
>
> Ideal conditions biostasis: This is where preservation procedures begin immediately after legal death is pronounced, typically with a standby team ready to initiate preservation procedures with minimal delay (usually less than 10 minutes after cardiac arrest). Under these conditions, brain structure is expected to remain largely intact and successful mechanical perfusion of the brain with preservative chemicals is expected to be possible.

> Less-than-ideal conditions biostasis: These can range from delayed pronouncement of legal death (more than 10 minutes after circulatory arrest), absence of a standby team (hours of delay), delayed discovery (days), to more complex situations involving autopsy or other complications. Under these conditions, brain structures will have progressively broken down, and mechanical perfusion will be increasingly difficult. The possibility for eventual recovery becomes progressively weaker as delay times increase.
>
> Which of these areas is your work primarily focused on?

Of the 20 respondents who chose one of these three options, 8 (40%) reported they focus primarily on ideal conditions, 4 (20%) on less-than-ideal-conditions, and 8 (40%) on both types roughly equally. Notably, what is called "ideal conditions biostasis" in humans is only "ideal" under today's law, in all jurisdictions that we are aware of. The highest quality preservation approach would be to start the procedure prior to the declaration of legal death, as is sometimes done for pets, or is routinely done for laboratory animals in neuroscience experiments. Of course, this raises ethical questions that are beyond the scope of this manuscript (McKenzie et al., 2024b).

**Assessing preservation quality**

*Memory and neural structure*

We asked respondents the following:

> Some neuroscientists have suggested that, while molecular and subcellular details play a role, the majority of information for long-term memories is likely physically stored in the brain at the level of neuronal connectivity patterns and ensembles of synaptic strengths (e.g. (Poo et al., 2016)).
>
> To what extent do you agree with the following statement: "The structural basis of long-term memories primarily consists of lasting changes in neuronal connectivity and ensembles of synaptic strengths, rather than in molecular or subcellular details."

Of the 21 respondents who picked one of the standard options, 11 (52%) answered "Agree" and 6 (29%) answered "Strongly agree", while 3 (14%) answered "Unsure" or "Neither agree nor disagree", and 1 answered "Disagree". The other respondent noted that they agreed that "long term memories primarily consist of lasting changes in neuronal connectivity and ensembles of synaptic strengths, but molecular/subcellular details should not be ignored." Taken together, this data suggests that most of the biostasis research community at least tentatively endorses the connectivity-based view of memory encoding (81% agreement), albeit with many not being particularly confident of this, and a significant minority being unsure or disagreeing.

*Surrogate biomarkers*

We asked respondents the following:

> A key question is which surrogate biomarkers have the potential to independently verify successful brain preservation for biostasis. Which of the following biomarkers, if measured in an adequate number of sampled parts of the brain, would convince you that the information critical for personal identity and most long-term memories has most likely been preserved (>95% confidence)? (Select all that you believe could meet your threshold for confidence).

The surrogate biomarker of neural connectivity as measured by ultrastructural imaging (e.g., connectome traceability) was selected by the majority of respondents (17, or 81%), making it the biomarker most widely accepted as dispositive of information preservation (**Figure 2**). Spatial biomolecular measurements (e.g., spatial proteomics, spatial transcriptomics) was endorsed by 9 (43%) respondents, while electrophysiological measurements (e.g., EEG recordings) was also selected by 9 (43%) as well. Cell viability measurements (e.g., Na+/K+ ratio) was chosen by 5 (24%) respondents. Notably, 4 (19%) respondents indicated that no surrogate biomarkers could convince them of successful information preservation.

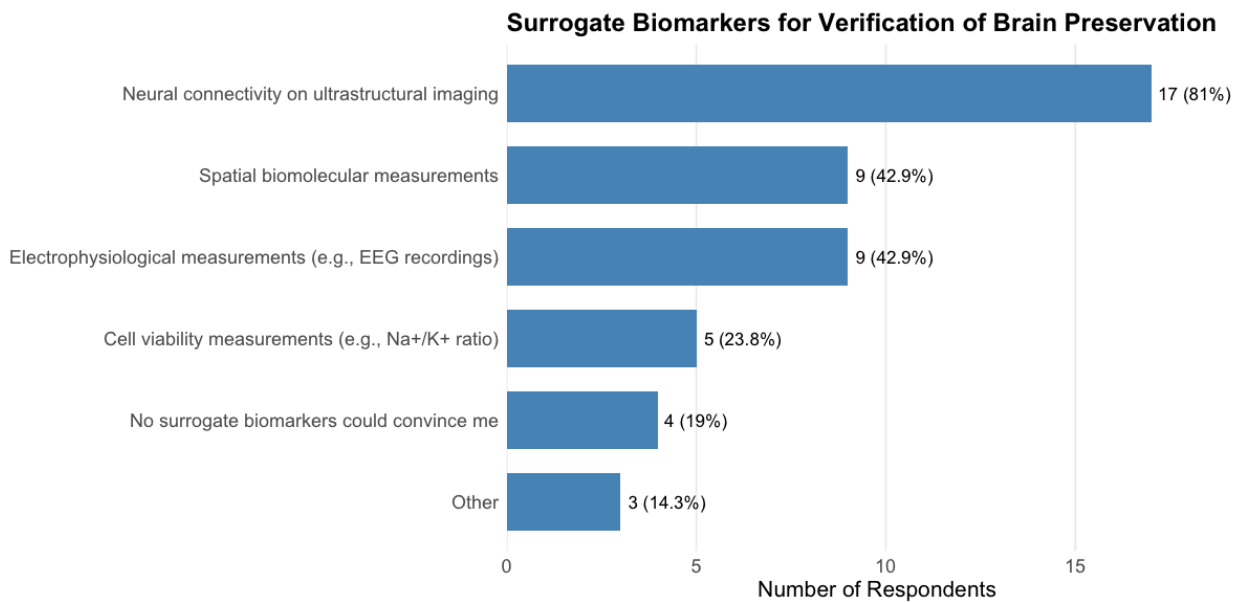

**Figure 2. Surrogate biomarkers for verification of brain preservation**. Respondents were invited to select multiple options if they would find multiple of them convincing that the information critical for personal identity and most long-term memories has most likely been preserved.

Three respondents provided additional written comments, which are noted in the "Other" category in the chart. One clarified that while they considered electrophysiological measurements, they would require "detailed single-unit-level recordings with dynamics that looked like they were recorded from the pre-preservation-state brain" rather than the more coarsely grained EEG recordings. Another emphasized the need for experimental validation, stating they would be convinced only after experiments showing "that preservation of neural connectivity correlates with preservation of long-term memories in mammalian brain." A third respondent suggested additional molecular markers including

"information on the location, quantity, and types of receptors (e.g., mGluRs) on neurons, as well as details about synaptic strength and classification (excitatory vs. inhibitory)."

Another possible biomarker brought up during discussion of this question is the structural integrity of neurons, independently of their viability. This was not included as one of the options here but could be included in a future iteration of this questionnaire. Notably, if synapses and neurite traceability is preserved, then one could argue that the 3D structure of neurons is likely largely preserved, although these could also be subtly decoupled, and requires experimental verification.

We also asked respondents the following:

> In what year do you think that the FDA or a similar organization in the US or another large country will first approve a biostasis procedure for medical use on the basis of a surrogate biomarker?

Responses revealed considerable uncertainty about this regulatory milestone, with 8 (36%) practitioners selecting "Unsure/don't know" (**Figure 3**). Among those who made predictions, the most common timeframe was 2060-2100 (5, or 23%), meaning that most practitioners believe such regulatory approval remains at least 35 years away. A smaller but notable group predicted earlier approval in the 2025-2040 (3 or 14%) or 2040-2060 (2 or 9%) windows, while another 2 (9%) pushed the timeline to 2100-2200. Two respondents (9%) predicted that such regulatory approval would never occur.

We note that this represents uncharted territory, as no biostasis procedure has ever received regulatory approval as a medical therapy, or come close to it. However, the concept seems to align well with regulatory trends. Agencies like the FDA increasingly accept surrogate endpoints when primary clinical outcomes would require impractically long follow-up periods. If stretched further, this approach could potentially be suitable for biostasis, where definitive outcomes might not be testable for decades or centuries. Of course, it would also require evidence that revival might one day be possible based on the preservation of the surrogate biomarker, as well as a forward-thinking regulatory agency. The wide distribution of opinions is indicative of the novelty of using surrogate biomarkers to assess a biostasis procedure (i.e., it has never been done before), as well as the unpredictable nature of future regulatory frameworks.

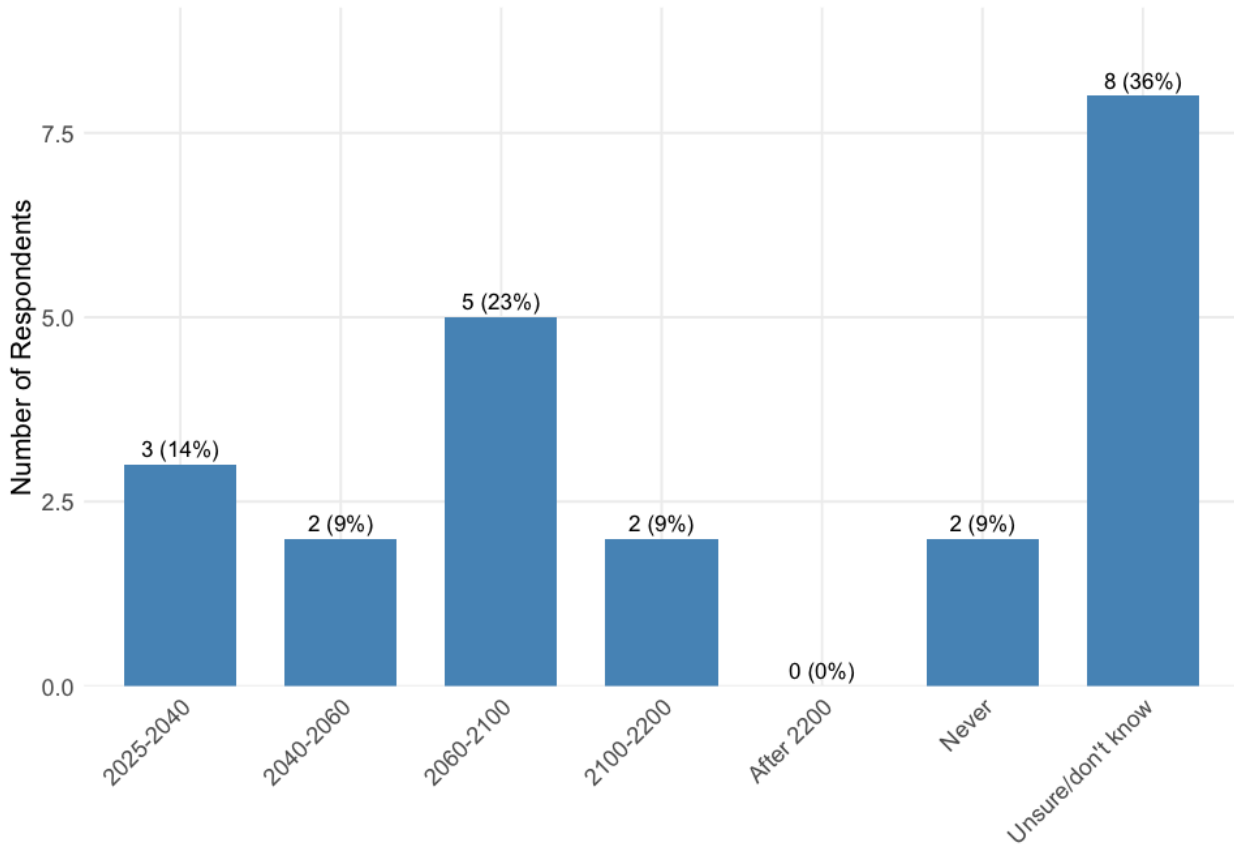

**Figure 3. Expected timeline for FDA approval of biostasis procedure.** Respondents estimated when the FDA or a similar organization will first approve a biostasis procedure for medical use based on surrogate biomarkers.

*Ischemic window for preservation*

For any organ, cells begin to deteriorate after blood flow stops. This very much includes the brain. After a long enough period of time, the decomposition of cellular structures is expected to become so severe that no future technology could reconstruct them, the same way we cannot decipher certain archaic writing systems like Linear A – i.e. there just is not enough available data. We asked respondents the following:

> Brain cells deteriorate after blood flow stops, the length of which is called the ischemic time. Eventually, they are expected to reach a point where no future technology could reconstruct the neural connections.
>
> In your estimation, how long after cardiac arrest would it become impossible to trace at least 95% of synapses to their originating cells, even with advanced future imaging and inference technologies? Please assume that external cooling via ice or refrigeration begins within minutes after the cessation of blood flow to the brain. And although there will also be biological variability in the speed of brain degradation, please assume an average case. Also, do not consider how difficult it may be to perfuse a brain after this time period, just consider the structure of the cells themselves.

Please indicate your best estimate about the time range when neural tracing becomes impossible for at least 95% of synapses:

- Check one box in the 50% column to indicate the time range where you believe there's a 50% chance this point occurs
- Check one box in the 99% column to indicate the time range where you're almost certain (99% confident) this point has occurred by.

Respondents revealed substantial variation in their estimated windows wherein measuring traceable synaptic connectivity would still be theoretically possible (**Figure 4**). At the 50% confidence level, most respondents believed neural tracing would become impossible somewhere between the windows of 3-6 hours and 2-5 days after cardiac arrest, with the steepest rise in the cumulative distribution occurring in the 12-24 hour window. At the 99% confidence level (near certainty), practitioners extended their estimates, with most believing irreversible damage would occur between 1-2 days and 5-10 days post-arrest. Notably, several respondents cited one paper as particularly useful in making this estimate (de Wolf et al., 2020). This study used electron microscopy to characterize the breakdown of cellular structures in rat brains after various intervals of normothermic and cold ischemia (de Wolf et al., 2020).

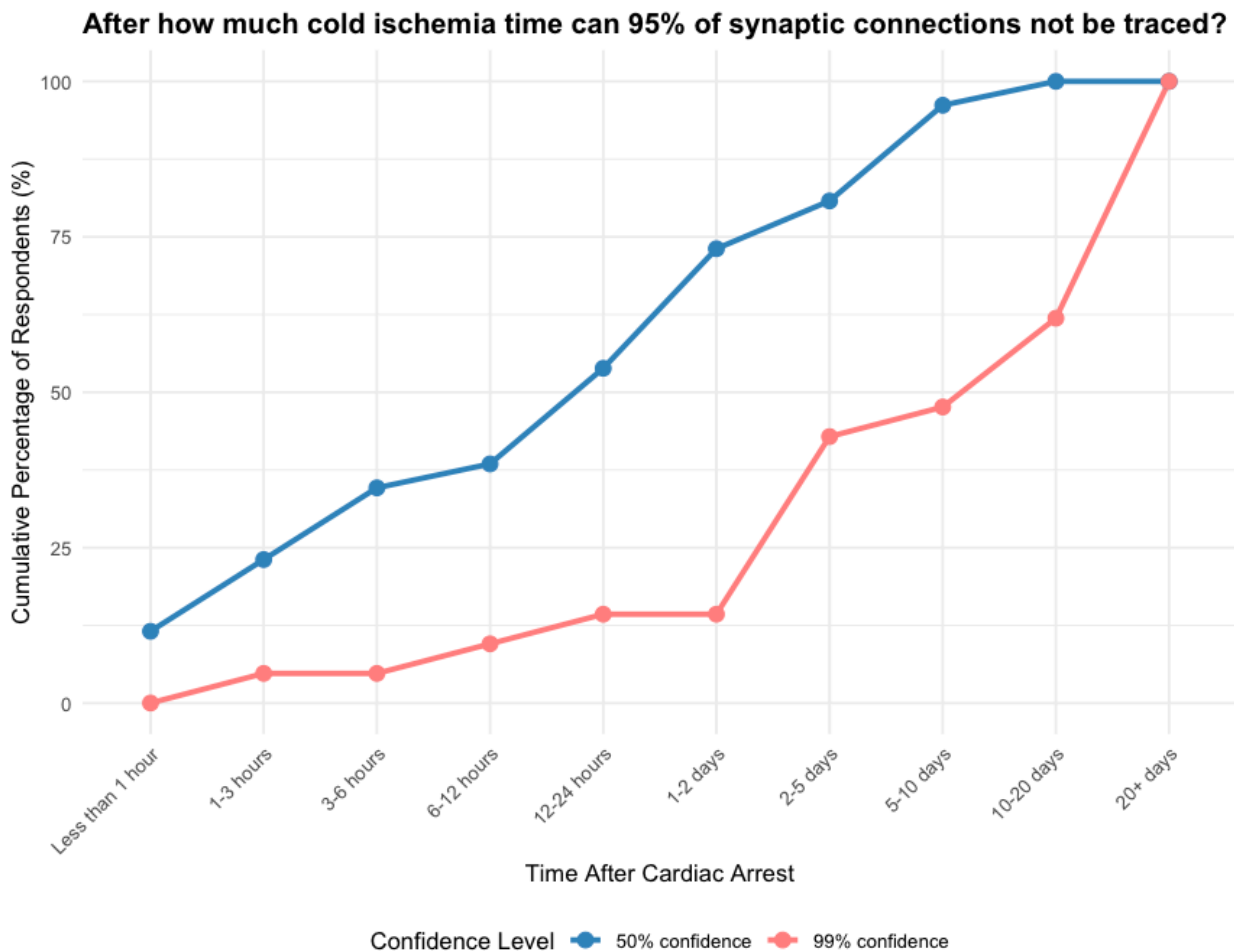

**Figure 4. Estimates of neural degradation timelines after cardiac arrest.** Cumulative distribution showing after what length of cold ischemia respondents expect that at least 95% of synaptic connections are no longer able to be traced, even assuming perfect inference technology, i.e. just based on the information still present. We asked respondents to estimate two confidence levels: 50% (blue) and 99% (red).

We also note that the threshold of being able to trace 95% of synapses to their originating cells is arbitrary. It is uncertain what percentage of synapses and associated connectivity information can be lost prior to information-theoretic death. And of course, this is on a spectrum, as it is possible that some memories but not all might be lost. Some data suggests significant redundancy of information processing to neuronal alterations. For example, one study found that removal of whole dendritic branches of V1 pyramidal neurons via micro-dissection – thus destroying thousands of synapses – did not significantly affect the orientation tuning of those neurons (Park et al., 2019). However, such redundancy studies are in their infancy, and the results depend on the context and type of memory studied. Further study of the degree of neural redundancy seems to be a critical question for contemporary biostasis, because some degree of damage is inevitable with ischemia, as well as other forms of damage associated with preservation.

Another problem with this question that was brought up in a discussion of this question is that the species of the animal was not specified. This might affect the rate of decomposition, due to differences in metabolic rates (Krassner et al., 2023).

One respondent brought up the point that some might incorrectly interpret this question as indicating the ischemic time window at which it is still worthwhile to perform a preservation procedure. But that is a separate question, which also depends on the method used for preservation, which in turn may depend critically on the ability to adequately perfuse the brain. And indeed, the perfusability of the brain also rapidly decreases following ischemia, which many believe occurs faster than the breakdown of cellular structure (Ames et al., 1968; Kloner et al., 2018; McFadden et al., 2019). One possible approach to this in the future is that there could be a separate question on the perfusability of the brain after different periods of ischemia. However, there is even less empirical data available to help respondents answer that question. The effects of ischemia on perfusion of the brain clearly requires more study. In the meantime, while these estimates can serve as a potential upper bound for the ischemic window in which successful preservation could be performed, they should not be interpreted as directly indicating when preservation procedures are still worthwhile, which depends on additional factors like perfusability.

**Most problematic aspects of contemporary biostasis**

For individuals signed up for biostasis, there are numerous potential obstacles that could prevent successful revival (Harris, 1989; More, 2023; Norton, 2020). We asked respondents to estimate the probability of various challenges interfering with successful preservation and eventual revival. To make this more practical, we asked respondents to consider a specific case. We asked respondents the following:

> If you are signed up for biostasis, and you legally die within the next month, this section asks: What are the reasons you expect that you personally will not be revived in the future? Or if you prefer to not discuss your own case, or you yourself are not signed up, imagine one specific other person you know who is signed up and whether biostasis will actually work for them if they legally die within the next month.
>
> For each option, please indicate your estimate of the probability that this potential problem on its own would lead to your information-theoretic death, conditional on everything else working out (meaning that the probabilities can certainly add to more than 100%). Please note that of course, the probabilities are not independent and it would not be appropriate to multiply them. For each, please answer on a percentage scale (0-100%).

This approach was designed to elicit concrete thinking about the practical challenges with modern biostasis techniques rather than hypothetical future scenarios. For each potential obstacle or "failure mode," respondents provided their estimate of the probability that this issue alone would cause information-theoretic death – i.e. the state where the information necessary for revival has been irretrievably lost (Merkle, 1992) (**Figure 5**).

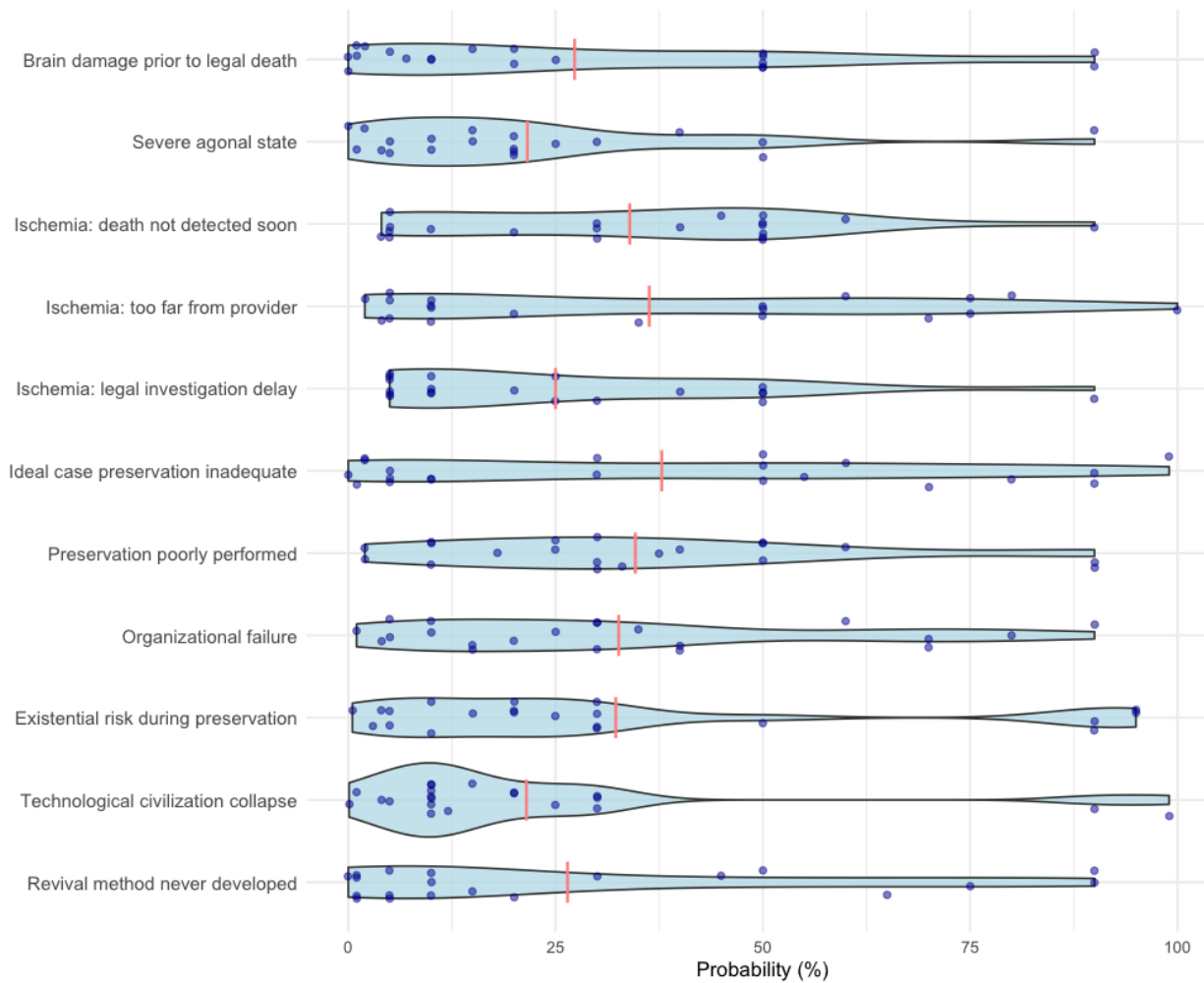

**Figure 5. Estimates of failure mode probabilities in contemporary biostasis.** Violin plots showing the distribution of estimates for the probability of different problems that could lead to information-theoretic death and therefore would prevent successful revival, also known as a potential "failure mode". Each blue dot shows an individual probability estimate, with vertical jittering (random position changes) for visualization clarity. Vertical red bars indicate the arithmetic mean of the probability estimate for each potential problem.

The biggest obstacle for success in biostasis identified among practitioners was that the ideal case preservation would be inadequate. Our respondents estimated a mean probability of 38% that even under optimal conditions with perfect execution, current preservation methods would be insufficient to prevent information-theoretic death. Closely following this was the probability of too long of an ischemic interval because of legal death occurring too far away from the biostasis provider, which received a mean estimate of 36%. This may be because there are very few biostasis providers, and our respondents may live too far away from all of them or the biostasis provider that they most favor. Independently of whether the preservation procedure could work in ideal circumstances, the preservation procedure being executed too poorly in practice was rated as the third-highest risk at 35%. This highly rated concern is not surprising, as there is limited professionalization in the field and little feedback about preservation efficacy in most cases. Partially this is a chicken-and-egg problem, because without preservation cases and the funding to perform them, it can be difficult for biostasis clinicians to get enough practice to acquire and maintain procedural skills.

The fourth most likely potential failure mode was too long of an ischemic interval due to legal death not being detected soon enough, with a mean probability estimate of 34%. This risk encompasses scenarios such as legal death at home during sleep, where hours might pass before cardiac arrest is noticed, or even days if the person lives alone and there is no check-in mechanism for them. Organizational failure during long-term preservation received a mean probability estimate of 33%, suggesting relatively high concerns about the institutional sustainability required for the decades or centuries of long-term preservation. Existential risk during preservation was estimated at 32%, representing the probability that all humans could die during the preservation period, thus making revival impossible regardless of technological progress. In other words, this estimate suggests that practitioners on average believe there is roughly a one-in-three chance that human extinction events could occur during the potentially centuries-long period of preservation required before people might ever be revived from biostasis, albeit with a wide distribution. Notably, revival technology could potentially benefit from the development of advanced AI technologies that could theoretically also pose an existential threat to humans. On the other hand, the more widespread use of biostasis could help to mitigate the significant time pressure that some feel to develop such technologies rapidly prior to death due to biological aging.

The remaining failure modes received somewhat lower but still substantial risk estimates. Brain damage prior to legal death averaged 27%, reflecting concerns about strokes, neurobiological disorders, or other brain injuries that could compromise the information content that preservation seeks to maintain. A revival method never being developed even though it is theoretically possible had a mean probability of 26%, which could reflect multiple concerns, including there being insufficient funding available to support its development. The ischemic interval being too long due to medicolegal death investigation received a mean estimate of 25%, indicating moderate concern about delays caused by autopsy requirements, coroner investigations, or other legal processes that could prevent timely preservation (McKenzie et al., 2025).

Too severe of an agonal state, based on different terminal conditions and the complications associated with the legal death process, could potentially compromise the preservation process to such an extent that the person no longer survives. The respondents estimated this probability at 22%. Historically, the largest concern with the agonal state has tended to be that it can lead to poor perfusion quality, for example, due to extensive thrombosis or hypoperfusion of the brain, which many believe is essential for high-quality preservation. It is often thought that this is poorly modeled in laboratory animals. Finally, the collapse of technological civilization received the lowest mean estimate at 22%, indicating that respondents are on average more concerned about the outright extinction of humanity than human civilization losing the capacity to perform long-term preservation and progress towards revival research.

Several respondents had nuanced comments about this probability assessment framework. One respondent clarified that they were answering each question as the probability of causing destruction of at least 50% of their mind, noting that "death is not all or nothing" and that their probability estimates of failure tended to be on the lower end as a result. The possibility of revival with personal identity partially intact is challenging to formulate precisely, but does seem to be a crucial consideration for future iterations of this set of questions, as this aspect of the question was unspecified (More, 1995 Chapter 2, "The Terminus of the Self"). Another respondent questioned the concreteness of information-theoretic death itself, suggesting that even without preservation of the brain, but perhaps a cell sample only, a clone created with detailed records of a person's experiences, hopes, dreams, photos, and genome might constitute a form of revival, asking "Have you got the person back if you do that?" This highlights a philosophical uncertainty about what constitutes successful revival and continuity of personal identity, which is also discussed below. While this is an interesting idea and might be very valuable to some people, this is generally not the purpose of biostasis. In any future versions of this questionnaire, it may be useful to clarify that the purpose of biostasis is the revival of a person on the basis of the information in their physical brain and possibly also the rest of their body, not a clone based on their genome and any artifacts they have generated (McKenzie et al., 2024b).

Several respondents had additional comments on these questions. Multiple respondents expressed concerns about AI-related existential risks, with one noting they "expect AI-related crises to develop faster than revival/uploading progress" and anticipating "AI dominance, a scenario which I do not expect to include revival." Others highlighted sociopolitical risks, including the possibility of government seizure of assets of biostasis organizations, regulatory shutdowns, anti-biostasis terrorism, and legal restrictions on whole brain emulation. One respondent noted that "society is generally predisposed to view claims regarding the potential for reversible human biostasis in a negative light" due to "deeply embedded cultural factors." Economic concerns were also raised, with one practitioner worried about revival remaining "too expensive given the amount provided for it for so long that I am eventually forgotten," particularly if energy costs remain high due to poor policy decisions. They estimated a 15% probability of this failure mode occurring. This illustrates one way that even successful preservation might not guarantee eventual revival if economic or social priorities shift over the extended time periods required for long-term care.

Overall, one take away from this exercise is that many respondents feel that many or all of these potential problems have a reasonable chance to lead to failure, but there is no consensus on problems that all agree will lead to a high risk of failure. We also note that these probability estimates are clearly highly correlated. For example, the underlying variable of how hard a problem revival of people from biostasis turns out to be in the future affects all of them. As a result, these probabilities cannot be multiplied to achieve a number for the overall probability of success.

**Preservation methods**

*Perfusion route*

The route of perfusion used to deliver preservation solutions to the brain is obviously a critical factor affecting the quality of perfusion, with numerous possible approaches that have been used in the literature and in cryonics (McFadden et al., 2019; McKenzie et al., 2024b). Two approaches that are commonly discussed are aortic or transcardiac cannulation and carotid ± vertebral cannulation, each with distinct theoretical advantages and practical considerations.

We asked respondents two questions. First:

> Which route of perfusion do you expect is most likely to lead to the most perfusion of the brain as rapidly as possible in an ideal case with minimal agonal phase or ischemia?

And second:

> Which route of perfusion do you expect is most likely to lead to the most perfusion of the brain as rapidly as possible in a non-ideal case, with 2 hours of warm ischemia and 10 hours of cold ischemia prior to initiating perfusion?

These questions take into account the speed of the procedure, the extent of the brain that will be perfused, and the reliability of the procedure in real-world cases. One refers to a case with ideal conditions and the other to one specific type of non-ideal conditions.

Under ideal conditions, the opinions of respondents were somewhat divided but showed a preference for aortic/transcardiac approaches, with 9 (41%) respondents either strongly favoring or leaning toward this method compared to 5 (23%) for carotid with or without vertebral cannulation (**Figure 6**). Under non-ideal conditions, preferences shifted, with 6 (27%) favoring aortic/transcardiac and 5 (23%) favoring carotid ± vertebral approaches, and a higher percentage of people being unsure (which was 4 or 18% of the respondents for ideal conditions and 8 or 36% for non-ideal conditions).

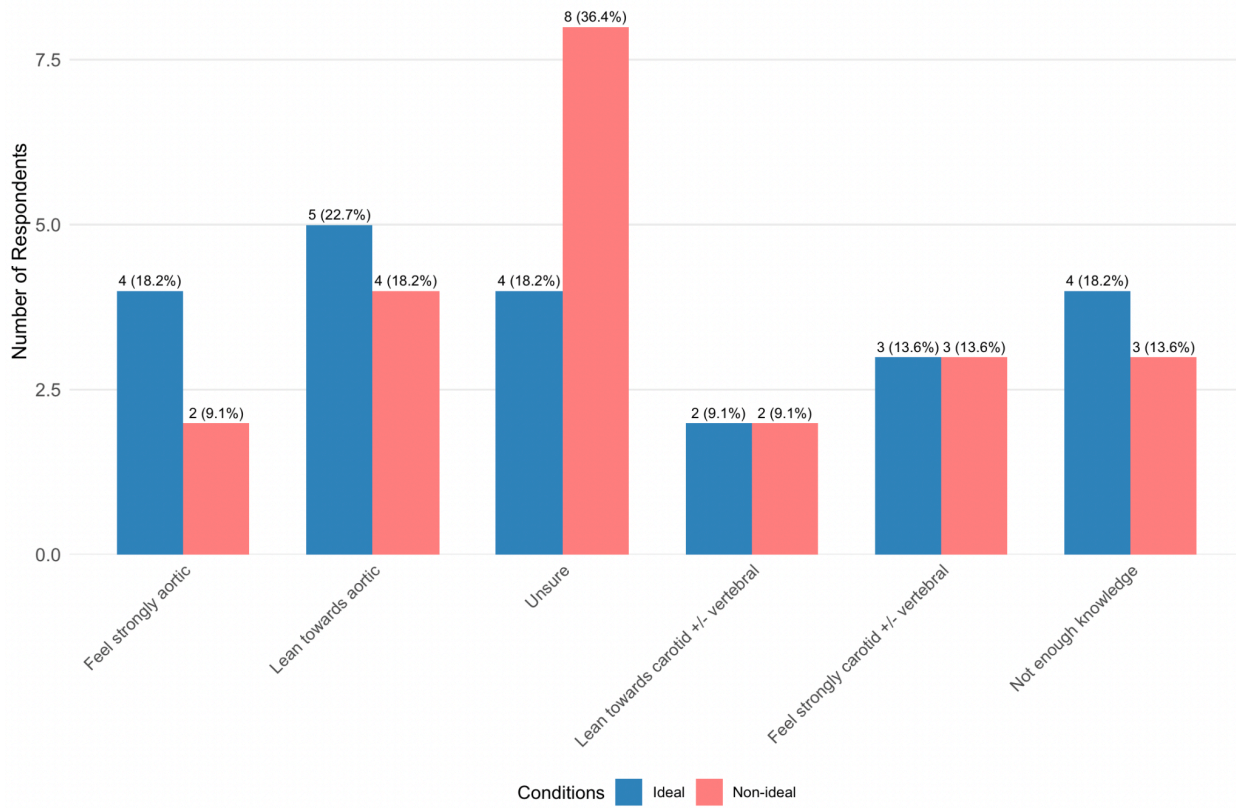

**Figure 6. Views on the most reliable perfusion route in ideal or non-ideal biostasis conditions.** Bar plots show the number (and percentage) of respondents selecting whether they view the aortic perfusate route as the best (either feeling strongly or leaning towards this), the carotid with or without vertebral route as the best, or are unsure of which is the best approach. This is shown for ideal conditions (blue) or non-ideal conditions, which is defined as 2 hours of warm ischemia and 10 hours of cold ischemia prior to the perfusion procedure (red).

One potential rationale for a slight divergence in the best approach for ideal compared to non-ideal conditions is that prolonged ischemia would cause more perfusion impairment. If perfusate preferentially flows to vasculature areas with less perfusion impairment, and the development of perfusion impairment is stochastic or even biased to occur more rapidly in the brain, then this might make cannulation of the arteries directly supplying the brain a relatively more reliable option in the setting of perfusion impairment. Perhaps more importantly, perfusion outcomes in non-ideal settings are not as well studied in general, which may have made respondents reluctant to be as confident.

A substantial proportion of respondents noted that they were unsure or expressed that they had insufficient knowledge to make a confident assessment. Further experimental testing, ideally directly comparing the two approaches, in both ideal and non-ideal conditions, would be the best approach to resolving this question definitively. Additionally, there may be differences between humans and other animals in terms of the best perfusion routes, as the body size, ease of surgical access, vascular anatomy, and the rate at which perfusion impairment develops has the potential to vary substantially between species.

*Views on advances in cryopreservation methods*

As summarized in the biostasis roadmap (McKenzie et al., 2024b), previous research in cryobiology has demonstrated that the pressure-driven introduction of cold liquids (such as silicones or fluorocarbons) or gases (such as oxygen, air, nitrogen, and helium) can accelerate cooling. After a brief description, we asked respondents:

> With what likelihood do you think cold gas or liquid cooling will have a paradigm-shifting effect on human biostasis protocols by 2050, being considered by the research community (at that time) as having a significant increase in the probability of successful revival?

Our respondents were divided on this question. 2 (9%) respondents considered it very likely, 8 (36%) likely, 3 (14%) unlikely, 3 (14%) very unlikely, and 6 (27%) were unsure or did not feel that they had sufficient knowledge to answer.

In recent years, certain pretreatment approaches (such as hormesis or cytokine treatment) have been found to reduce cryoprotectant toxicity, potentially by mobilizing cellular defense systems (Mazur et al., 2024). After a brief description, we asked respondents:

> With what likelihood do you think pretreatment approaches for reducing cryoprotectant toxicity will have a paradigm-shifting effect on human biostasis protocols by 2050, being considered by the research community (at that time) as having a significant increase in the probability of successful revival?

3 (17%) respondents considered this very likely, 5 (23%) likely, 4 (18%) unlikely, 1 (4.5%) very unlikely, and 9 (41%) felt unsure or did not feel that they had sufficient knowledge to answer. Altogether, 36% of respondents could be considered optimistic or cautiously optimistic about this line of research paying significant dividends for cryopreservation approaches over the next 25 years.

The blood-brain barrier (BBB) presents a significant challenge for the delivery of certain preservation compounds to brain tissue. Various approaches have been proposed to modify the BBB, including chemical methods (e.g., sodium dodecyl sulfate) and ultrasound techniques, as summarized in our biostasis roadmap article (McKenzie et al., 2024b). The use of detergents is controversial in part because of their potential effect on the lipidome, which is uncertain (Osetrova et al., 2024). Notably, BBB modifiers are sometimes not considered as important for preservation methods perfusing only fixatives, as aldehyde preservatives such as formaldehyde can cross the BBB even in the absence of BBB modifiers (Shcherbakova et al., 1986). We asked respondents:

> In ideal cases with very minimal ischemia, which blood-brain barrier modification techniques do you believe are the best near-term option for improving preservation quality?

6 (27%) reported that they thought chemical approaches such as sodium dodecyl sulfate were the most promising option, 1 (4.5%) ultrasound techniques, and 2 (9%) combined approaches. The majority of respondents, 12 (55%), were unsure. Another possibility for introducing cryoprotective agents, which has recently been proposed, is to close the BBB for water by targeting brain osmoregulation via aquaporin 4 in astrocyte endfeet (German et al., 2025). This could be listed as another option in a subsequent questionnaire.

The temperature of long-term preservation is an important consideration. For pure cryopreservation approaches, liquid nitrogen storage at -196°C has been the standard practice, but intermediate temperature storage (ITS), typically between -140°C and -130°C, has been proposed as a potential alternative to minimize thermal stress and resulting fracturing of preserved tissues (Wowk, 2011). We asked respondents:

> For cryopreservation-only methods, which storage approach do you currently favor for use, based on the publicly available evidence, neglecting cost and assuming equivalent security?

16 (73%) respondents answered ITS, 3 (14%) answered traditional liquid nitrogen storage, and 3 (14%) were unsure. However, this question was clearly limited because it asks respondents to ignore core downsides of contemporary ITS systems. One respondent noted: "Given the relative lack of cracking in brains compared to other organs, I doubt the value of ITS. However, I [chose] ITS because you specified no difference in cost or security." In future questionnaires, this question should be better formulated to actually address the trade-offs or simply removed.

A major potential milestone for pure cryopreservation approaches would be the ability to reversibly cryopreserve the brain with global electrophysiological functions intact. Notably, this milestone could not be reached for fixative-based preservation methods, at least prior to the development of molecular nanotechnology or similarly precise chemical engineering, at which point it may be effectively a revival method. We asked respondents the following:

> By what year do you estimate that there will be a publication demonstrating the reversible cryopreservation of a mammalian brain with recovery of global electrophysiological functions, such as EEG recordings for more than one hour that are similar to those occurring during life?

Respondents tended to be quite optimistic about this milestone occurring within the next 15 years, with 60% responding that it had a 50% chance of happening by the 2030-2040 time window (**Figure 7**). Respondents had substantially longer timeline estimates when asked by when they thought this advance would occur with high probability. 80% of respondents gave a 90% probability estimate that such a finding would be published by 2100.

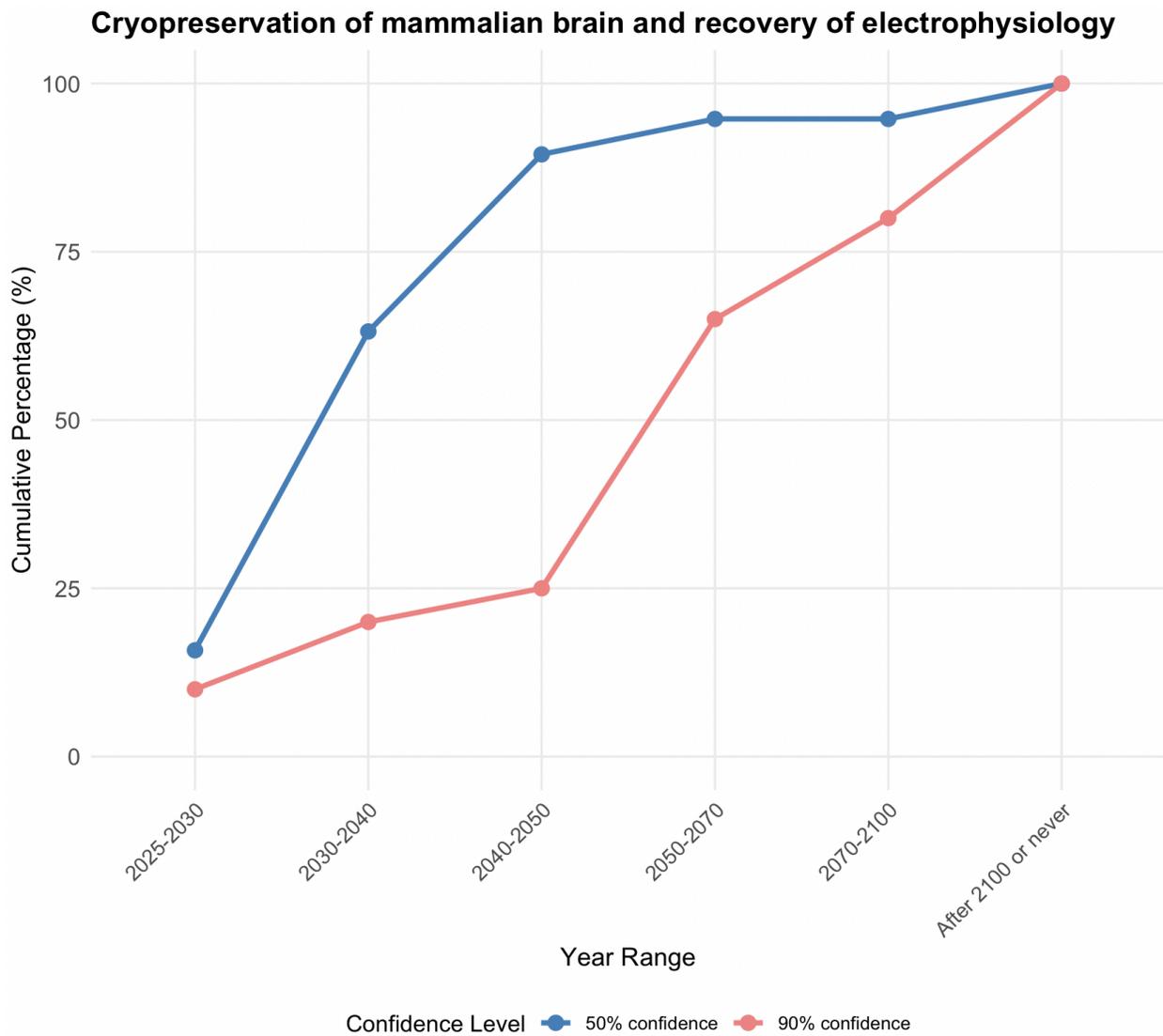

**Figure 7**. Timeline estimates for reversible cryopreservation of mammalian brain with recovery of global electrophysiology. Cumulative distribution showing practitioner estimates of when a publication will report the reversible cryopreservation of a mammalian brain with recovery of global electrophysiological functions, such as EEG recordings for >1 hour similar to those during life. Respondents provided timeline estimates at two confidence levels: 50% confidence (blue) and 90% confidence (red).

*Aldehyde preservation and long-term fluid preservation*

Immersion fixation of the brain is a method often used in neuropathology and brain banking for preserving the structure of the brain for future study (Garrood et al., 2025). It has also been used in biostasis as well. We asked respondents the following question:

> How likely do you think it is that immersion fixation alone, if performed under ideal circumstances (e.g. initiated just after legal death, refrigeration during the immersion) can preserve adequate brain structure to infer the synaptic connectivity across the entire human brain?

The opinions of our respondents varied on this question, but leaned skeptical, with 8 (36%) considering it unlikely or very unlikely, 5 (23%) considering it likely, 1 (4.5%) very likely, 5 (23%) unsure, and 2 (9%) reporting insufficient knowledge. One remaining respondent (representing 4.5% of the sample) wrote in a text-based response instead of a standard option: "'Across the entire human brain' - unsure how well immersion can preserve long-range subcortical structures, and whether the fixatives get there fast enough in a refrigeration case." This response reflects a common concern that deep brain regions will not be well enough preserved via immersion fixation, due to the length of time that it takes for fixative to reach these areas via diffusion.

An alternative to long-term preservation via liquid nitrogen is fluid preservation, where fixed brain tissue is stored in preservative solutions at refrigerator temperatures rather than being vitrified (McKenzie et al., 2024a). We asked respondents the following question:

> Assume that the synaptic connectivity of a brain is well preserved via perfusion-based aldehyde fixation very soon after legal death. How likely do you think it is that preservation of such a brain for 50 years in a solution containing aldehyde fixatives at refrigerator temperature (4°C) would maintain sufficient structural integrity to perform accurate tracing of synaptic connectivity throughout the entire brain?

Respondents were relatively more optimistic about this scenario. 8 (36%) considered it very likely, 8 (36%) likely, 4 (18%) were unsure, and 1 (4.5%) considered it unlikely, while 1 (4.5%) didn't feel knowledgeable enough to answer. This data suggests that while practitioners have concerns about the initial preservation quality achievable with immersion fixation alone, they are more confident in the long-term stability of brain tissue that has been successfully fixed with aldehydes via perfusion.

*Preservation of synaptic connectivity as seen via electron microscopy*

In the early 2010s, the Brain Preservation Foundation (BPF), a non-profit organization, organized a prize which would be given to the first organization to demonstrate the preservation of high-quality ultrastructural preservation across an entire brain, alongside a solid argument that the structure could be preserved for at least 100 years (Hayworth, 2011). The key metric was the ability of human annotators to trace neural processes from synapses to their originating neurites, and to continue to trace those neurites for extended distances without discontinuities across an entire embedded tissue block imaged via volume electron microscopy (Hayworth, 2011). Notably, unmyelinated axons are the most challenging type of neurite to trace in such a data set, as they tend to have a smaller diameter and greater length than dendrites (Januszewski et al., 2025).

The sampling strategy used for the BPF prize was to: (a) first, examine the brain macroscopically (e.g. via gross examination of sections and/or x-ray imaging) to identify any regions that looked like they may have poor preservation, which would then be biopsied for imaging, as well as (b) take additional biopsy samples from other arbitrarily or randomly sampled areas of the brain. The idea behind this sampling strategy is that it would provide both targeted assessment of potentially problematic areas, as well as systematic assessment of overall preservation quality throughout the brain.

The BPF prize was won fairly quickly by a team at 21st Century Medicine, who published their method, called aldehyde-stabilized cryopreservation (ASC), in 2015 (McIntyre and Fahy, 2015). The BPF Small Mammal Prize, which was tested on rabbit brains, was officially won in 2016. And the Large Mammal Prize, which was tested on pig brains, was officially won in 2018 (Wiley, 2018).

The co-founder of the BPF, Kenneth Hayworth, then called for the method to be developed into a medical procedure, with a documented quality control procedure (Hayworth, 2018). He noted that "[o]f course this will likely require some modifications to the published ASC procedure that has, as of this writing, only been tested on animals in ideal laboratory conditions." However, no method, including ASC, has yet been found to preserve this level of structural fidelity across an entire human brain. Additionally, there has not yet been a published result with similar proven structural brain preservation quality with any other preservation method.

This BPF prize is well-known in the biostasis community. We asked respondents to estimate when different types of preservation technologies would demonstrate the same capability in both animal models and human brains. We note that although the BPF tracing criteria is well-established, and we are focusing on it with these questions, we are not claiming that it is necessarily the best surrogate biomarker for the potential of biostasis to work, and reasonable people can certainly disagree about how dispositive it is of the potential for future revival.

First, we asked about when cryopreservation methods that do not use aldehyde fixatives would be reported in a publication to have the same level of neural connectivity preservation as ASC achieved in the BPF prize. For context, historically, when "pure" cryopreservation techniques are used to preserve entire brains (i.e., without aldehyde preservatives), measuring the neural ultrastructure has been challenging. In particular, osmotic shrinkage during the cryoprotectant perfusion process, which can reduce brain volume significantly, has made it difficult to clearly assess the preservation of synaptic connectivity ("21CM Cryopreservation Eval Page – The Brain Preservation Foundation," 2015). As a result, it has been difficult to clearly identify synapses and trace their associated neurites. And without the ability to directly trace neurites, it is possible that potential structural damage due to ischemia, rapid osmotic shifts, cryoprotective agent toxicity, or other factors could be masked. The specific question we asked was this:

> By what year do you estimate that there will be a publication, in prospective animal experiments, showing the preservation of whole-brain synaptic connectivity using cryopreservation methods without aldehydes? Specifically, this would require volumetric ultrastructural imaging showing the ability to continuously trace neural processes from synapses to their originating neurites for extended distances without discontinuities, in sampled regions throughout the entire brain, comparable to what was achieved in the Brain Preservation Foundation (BPF) prize won by 21st Century Medicine using aldehyde-stabilized cryopreservation.

Our respondents' predictions for achieving BPF-level synaptic connectivity preservation using pure cryopreservation methods showed a good degree of optimism in the near- to medium-term, with 10 (56%) respondents estimating this milestone would be reached by the 2030-2040 time window (**Figure 8**). At the 90% confidence level, practitioners were more conservative, with a notable proportion of 5 (25%) indicating they would not be 90% confident that this achievement would occur until either after 2100 or never.

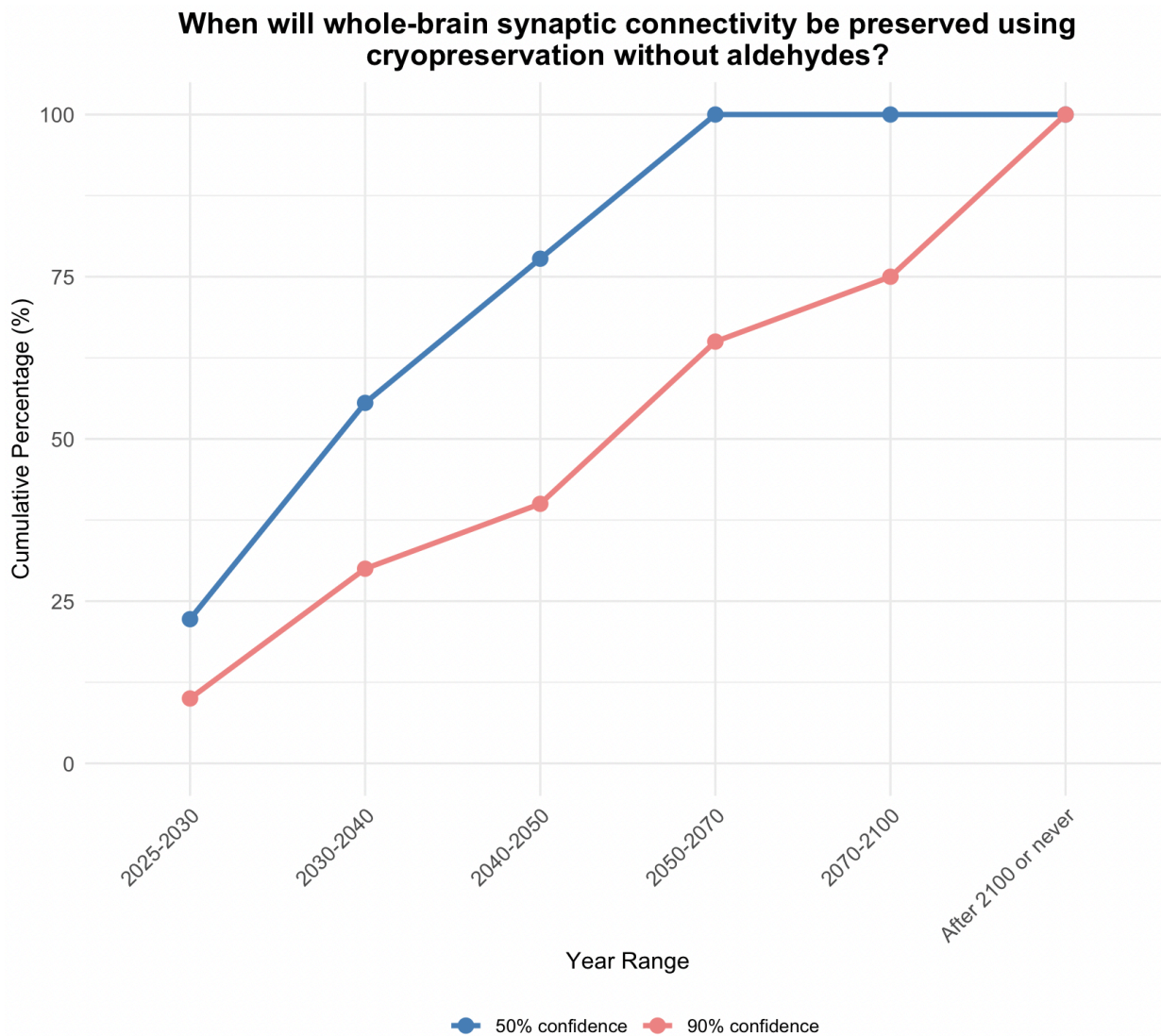

**Figure 8**. Timeline estimates for demonstrably achieving preservation of whole-brain synaptic connectivity using cryopreservation without aldehydes in animal experiments. Cumulative distribution showing when respondents estimate this milestone will be achieved at 50% (blue) and 90% (red) confidence levels.

Second, we asked the following question:

By what year do you estimate that there will be a publication, in multiple donated human brains, showing the preservation of whole-brain synaptic connectivity using cryopreservation methods without aldehydes? Same BPF tracing criteria as in the previous question.

As expected, practitioner predictions for this milestone were notably more conservative than for laboratory animals. The human brain poses multiple practical challenges, including the size of the organ relative to most laboratory animals used, as well as the inability to start the procedure prior to the onset of ischemia. However, it is clearly the most important organ for the field of human biostasis. At the 50% confidence level, practitioner estimates showed a gradual progression, with 7 (39%) respondents believing this would be achieved by 2030-2040, 12 (67%) by 2040-2050, and agreement among all 19 (100%) that this was 50% likely to occur by 2050-2070 (**Figure 9**). At the 90% confidence level, practitioners were considerably more cautious, with 11 (55%) respondents having this level of confidence that the milestone would be reported in a publication by the 2050-2070 time window.

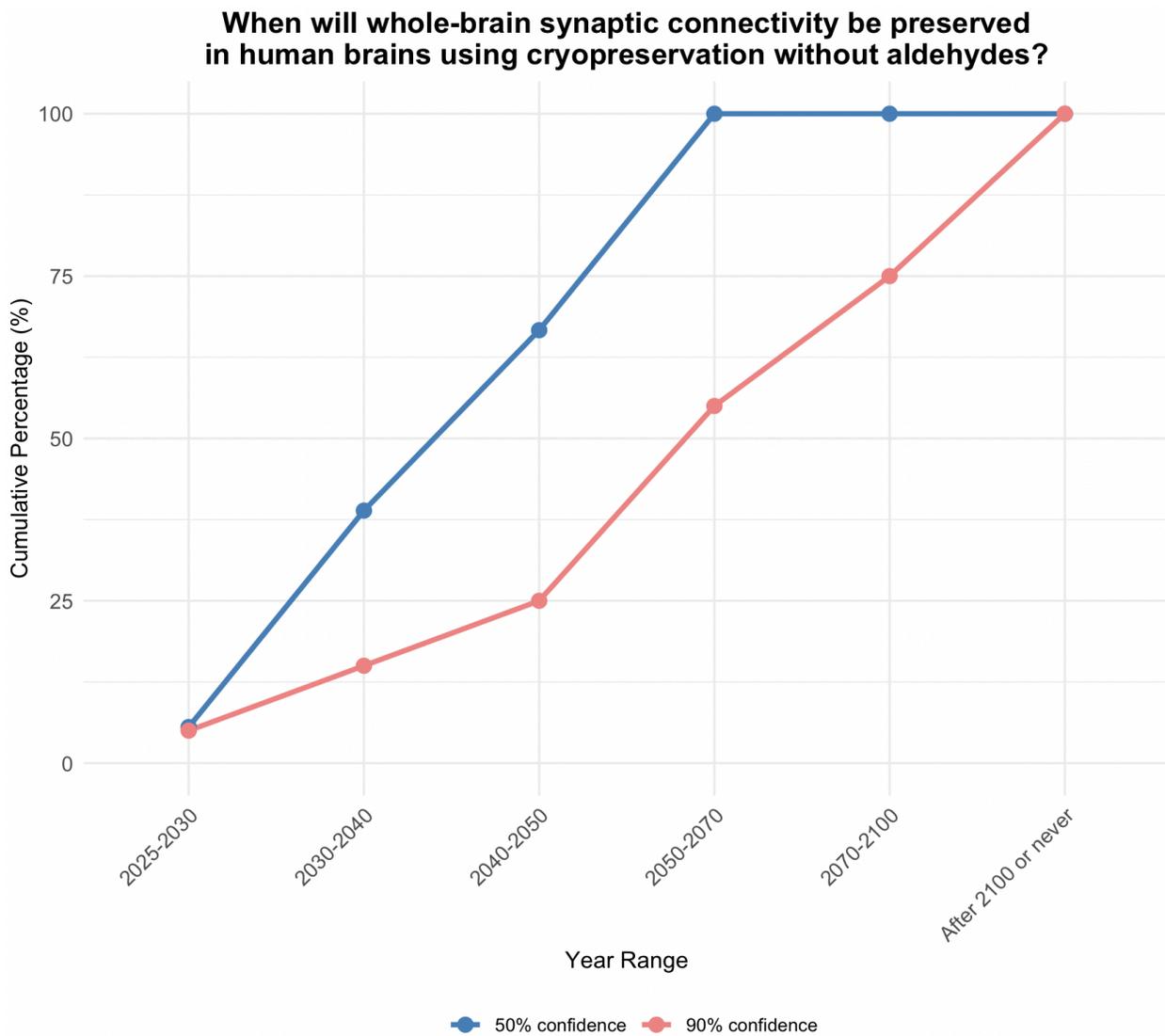

**Figure 9**. Timeline estimates for demonstrably achieving preservation of whole-brain synaptic connectivity using cryopreservation without aldehydes in human brains. Cumulative distribution

showing when respondents estimate this milestone will be achieved at 50% (blue) and 90% (red) confidence levels.

Third, we asked the following question:

> The BPF prize for aldehyde-stabilized cryopreservation (ASC) was awarded for preserving synaptic connectivity in laboratory animal brains (McIntyre and Fahy, 2015), but this has not yet been shown in human brains. By what year do you estimate that researchers will demonstrate, in multiple donated human brains, the preservation of whole-brain synaptic connectivity using ASC? Same BPF tracing criteria as the previous questions.

Practitioner predictions were more optimistic about this question, with 100% of respondents believing this milestone would be achieved by 2030-2040 at the 50% confidence level (**Figure 10**). At the 90% confidence level, practitioners still showed substantial optimism, with 15 (79%) expressing high confidence in this being achieved by the 2040-2050 window. Our respondents were likely more optimistic about these timeline estimates because aldehyde-based preservation has already demonstrated structural preservation in laboratory animals for the BPF prize. However, we note that there are considerable practical challenges necessary to be surmounted to achieve this in humans, most especially related to the difficulty of achieving high-quality perfusion across the ischemic human brain, or, alternatively, figuring out an approach for which high-quality perfusion is not essential. As a result, the achievement of this milestone should certainly not be considered an inevitability.

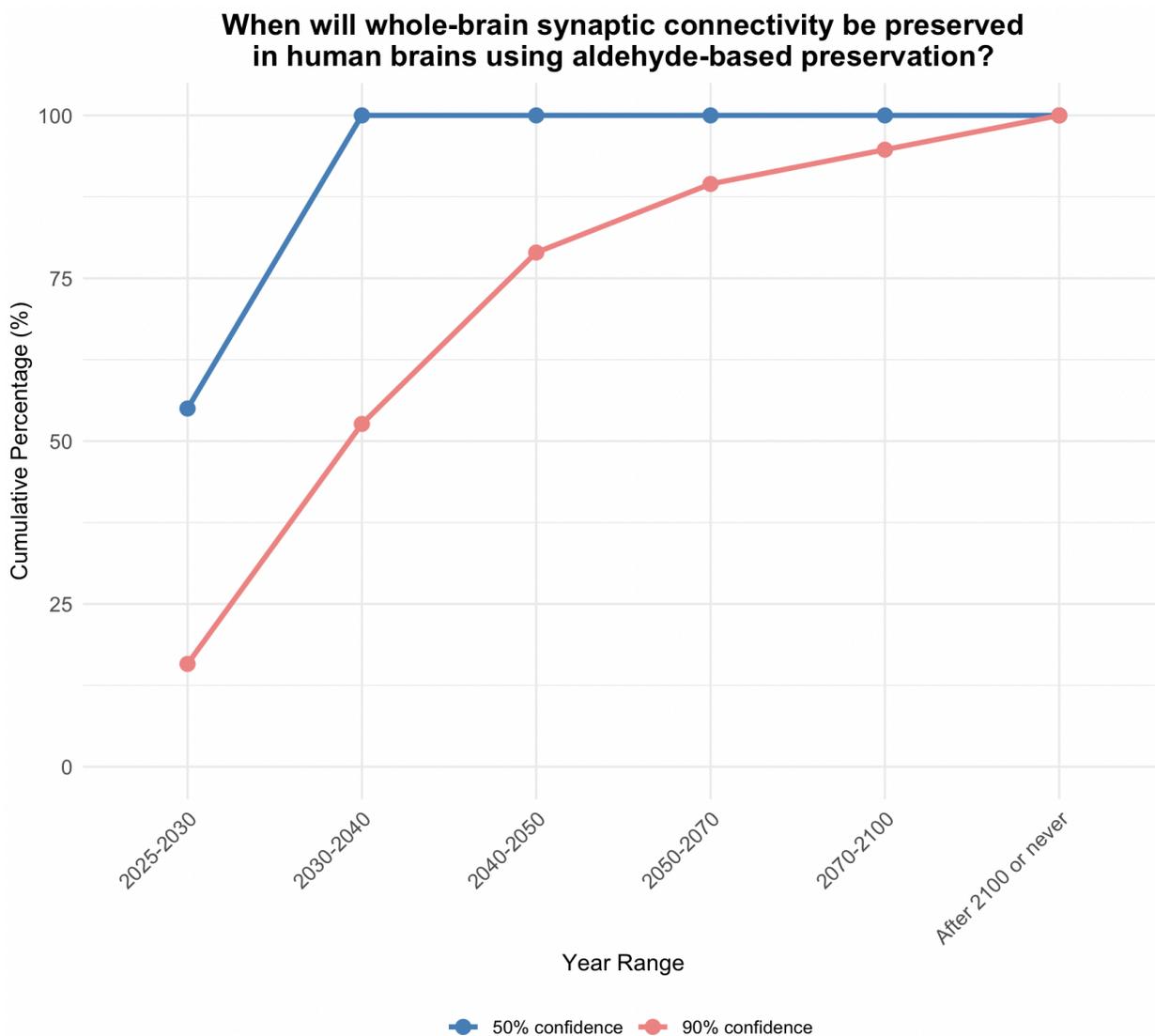

**Figure 10**. Timeline estimates for demonstrably achieving preservation of whole-brain synaptic connectivity using aldehyde-stabilized cryopreservation (ASC) in human brains. Cumulative distribution showing when respondents estimate this milestone will be achieved at 50% (blue) and 90% (red) confidence levels.

Just because something can be done once or a few times, does not mean that it can be reliably done. An additional future question might be to assess estimates of the reliability of this procedure under various types of real-world conditions. As a result, there will remain plenty of practical research and development work ahead to be performed on aldehyde-based preservation methods, even if this milestone is eventually reached.

**Provably reversible preservation timelines**

The most obviously game-changing milestone in biostasis research would be the demonstration of provably reversible preservation in whole organisms. Although reversible cryopreservation has been achieved in small organisms like nematodes (Vita-More and Barranco, 2015), this has not yet been demonstrated in mammals. While progress has been made on reversibly cryopreserving individual vital organs, cryopreserving all vital organs at once, as well as other body parts and body cavities necessary for healthy life, is clearly a Herculean task (Wowk, 2025). Some might argue it is a Sisyphean one, unless there is a paradigm shift in the entire approach to the problem. One of the key problems is that there are substantial differences in cryoprotectant absorption, toxicity responses, and ice formation sensitivities across tissue types (Wowk, 2025).

Aldehyde-based preservation methods could not achieve this milestone without the development of molecular nanotechnology or a similarly powerful future technology, as aldehydes create cross-links that current technology cannot undo in a targeted enough manner. As a result, we exclusively ask these forecasting questions about pure cryopreservation approaches to biostasis. We first asked respondents the following question:

> By when do you estimate there will be a publication with a credible description of provably reversible cryopreservation of a whole laboratory animal, including recovery to a long-lived healthy state?

(Notably, this question was imprecisely stated because it should have explicitly referred to mammalian revival, but we suspect that most respondents viewed it this way, due to previous context in the questionnaire, despite the lack of precise wording in the question text.)

Practitioner predictions for the reversible cryopreservation of a whole laboratory animal showed considerable optimism in the medium term, with 11 (55%) estimating that this milestone would be achieved by the 2040-2050 time window with 50% confidence (**Figure 11**). However, when respondents were asked for their 90% confidence estimates, their answers were considerably more conservative, with 10 (50%) believing there would be a 90% likelihood of success by the 2070-2100 time window, and 10 (50%) setting their high-confidence timeline at after 2100 or never. As a result, while many practitioners are cautiously optimistic about eventual success of this milestone, the uncertainty in these predictions suggests that it should not be relied upon to be achieved in this century, at least according to these estimates.

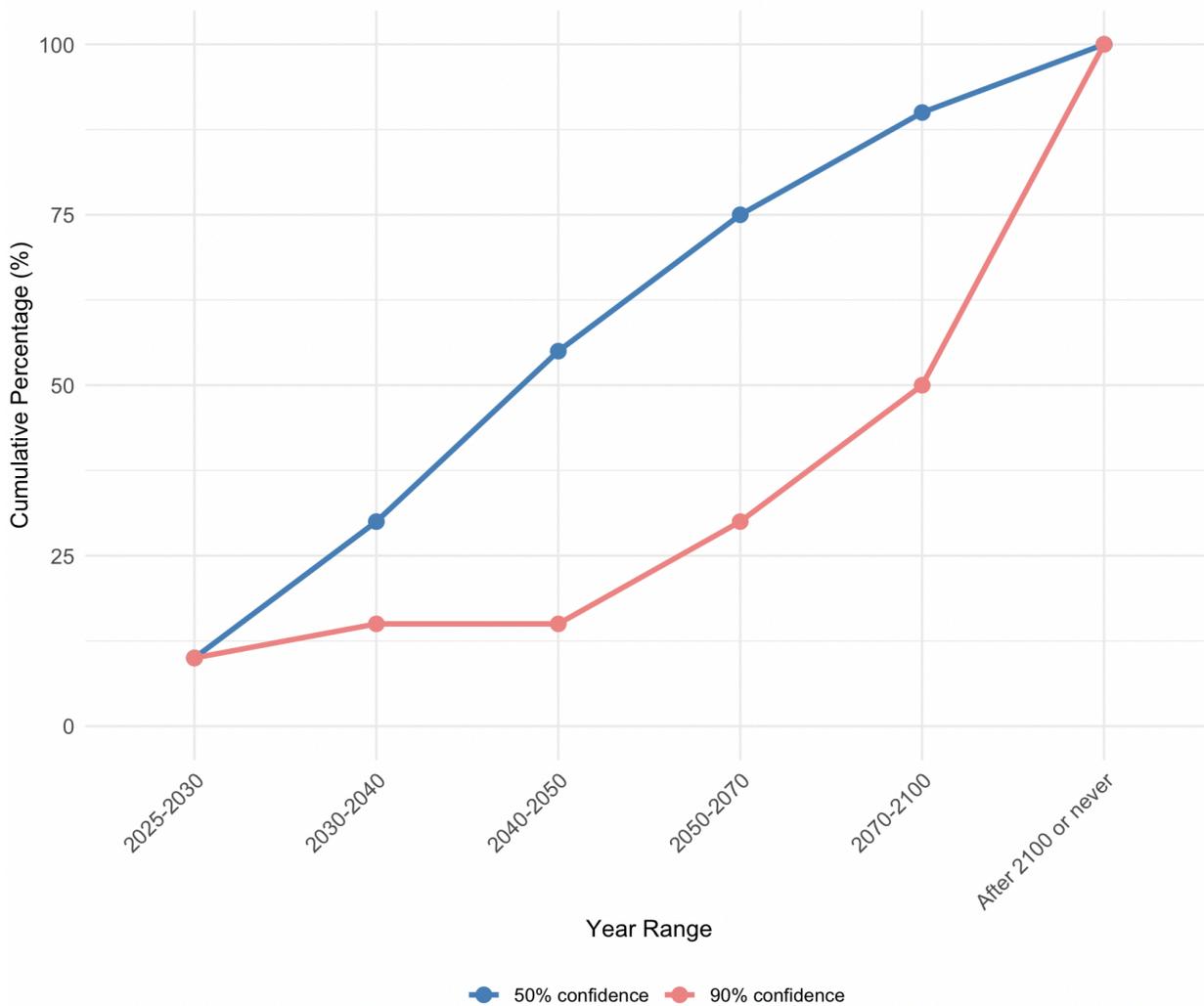

**Figure 11**. Timeline estimates for demonstrating reversible cryopreservation of whole laboratory animals. Cumulative distribution showing when respondents estimate a publication describing successful revival to a long-lived healthy state at 50% (blue) and 90% (red) confidence levels.

We then asked respondents the following:

> By when do you estimate there will be a publication with a credible description of provably reversible cryopreservation in a human patient, including recovery to a long-lived healthy state?

At the 50% confidence level, most respondents believed this milestone would be achieved by 2070-2100, with 15 (78.9%) estimating it would be achieved by that time window (**Figure 12**). When asked for 90% confidence estimates, the timeline extended further, with the clear majority of 14 (70%) practitioners reporting they could only be nearly certain of human reversible cryopreservation until after 2100, if ever. Taken together, a minority of respondents believe that this milestone is 50% likely to be achieved within the next 25 years, and a minority are 90% confident that it will be achieved in this century.

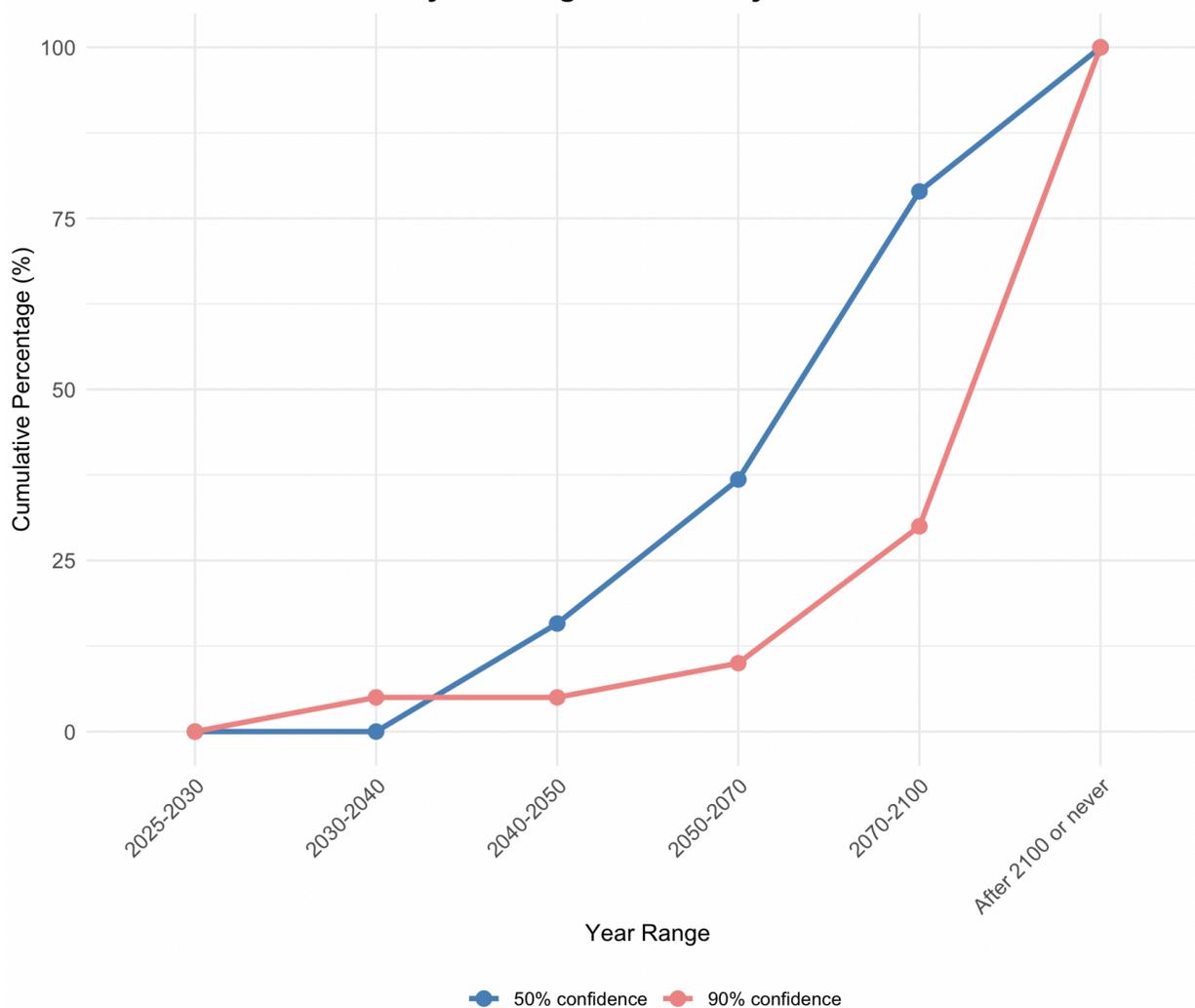

**Figure 12**. Timeline estimates for demonstrating reversible cryopreservation in human patients. Cumulative distribution showing when respondents estimate a publication describing successful revival of humans to a long-lived healthy state at 50% (blue) and 90% (red) confidence levels. One respondent provided only a 90% confidence estimate (at a relatively early year range) without a corresponding 50% estimate, causing the apparent illogical crossing of the lines.

**Revival methods after surrogate biomarker preservation**

*Preamble of revival methods*

Prior to asking questions about potential revival methods from current preservation procedures, we first gave the following preamble:

> We can consider three classes of revival technologies that have been proposed: conventional cell repair, reconstruction based on molecular nanotechnology, and whole brain emulation.

Conventional Cell Repair: This theoretical approach proposes using advanced medical technologies such as "nanobots" to repair preserved tissues at the cellular level (see Freitas 2022) (R. Freitas, 2022). Conventional cell repair would involve scanning the preserved body to map all relevant structures at subcellular resolution (~100 nm), then carefully warming the body enough to allow removal of potentially harmful metabolites while cells return to a liquid state. However, this approach faces challenges due to the molecular damage that occurs during preservation.

Molecular Nanotechnology: This theoretical approach involves using atomically precise machines to repair damage at the molecular level. It would theoretically allow for more comprehensive repair than "conventional cell repair", by allowing for the manipulation of individual atoms and molecules. While it seems to not break known laws of physics, this would require significant technological advances beyond current capabilities.

Whole Brain Emulation: The proposed procedure here generally involves extracting the detailed neural architecture of the preserved person's brain through scanning techniques, then constructing a functional computational representation of its information-processing system. Once the model for this "digital person" is created, the person's life could be restarted either within a computer simulation environment or via interface with physical systems through robotics. Many have philosophical questions about this approach, particularly regarding whether emulation would allow personal psychological experience to genuinely continue, or if the resulting entity would merely be a highly sophisticated "copy."

Obviously, there is a large amount of potential heterogeneity within potential ways that these methods could be implemented (see McKenzie et al. 2024) (McKenzie et al., 2024b).

Imagine at some point in the future that technology has advanced to the point that if these technologies will ever be practical, they are able to be developed. Also imagine that biostasis organizations have survived with patients maintained in preservation until then. Also put aside for the moment their desirability and other questions such as whether humanity and biostasis organizations will survive that long.

*Technical feasibility of different methods*

For each of these proposed technologies, we then asked respondents the following three questions:

What is your probability estimate that {conventional cell repair, molecular nanotechnology, whole brain emulation} will ever be technically able to revive any people who were preserved in 2025 and earlier? Please answer as a percentage (0-100%).

Respondents answered that whole brain emulation was the most likely to be technically able to revive people preserved in 2025 and earlier, with an average of 62% (**Figure 13**). Molecular nanotechnology was close at 53%, followed by conventional cell repair at 23%.

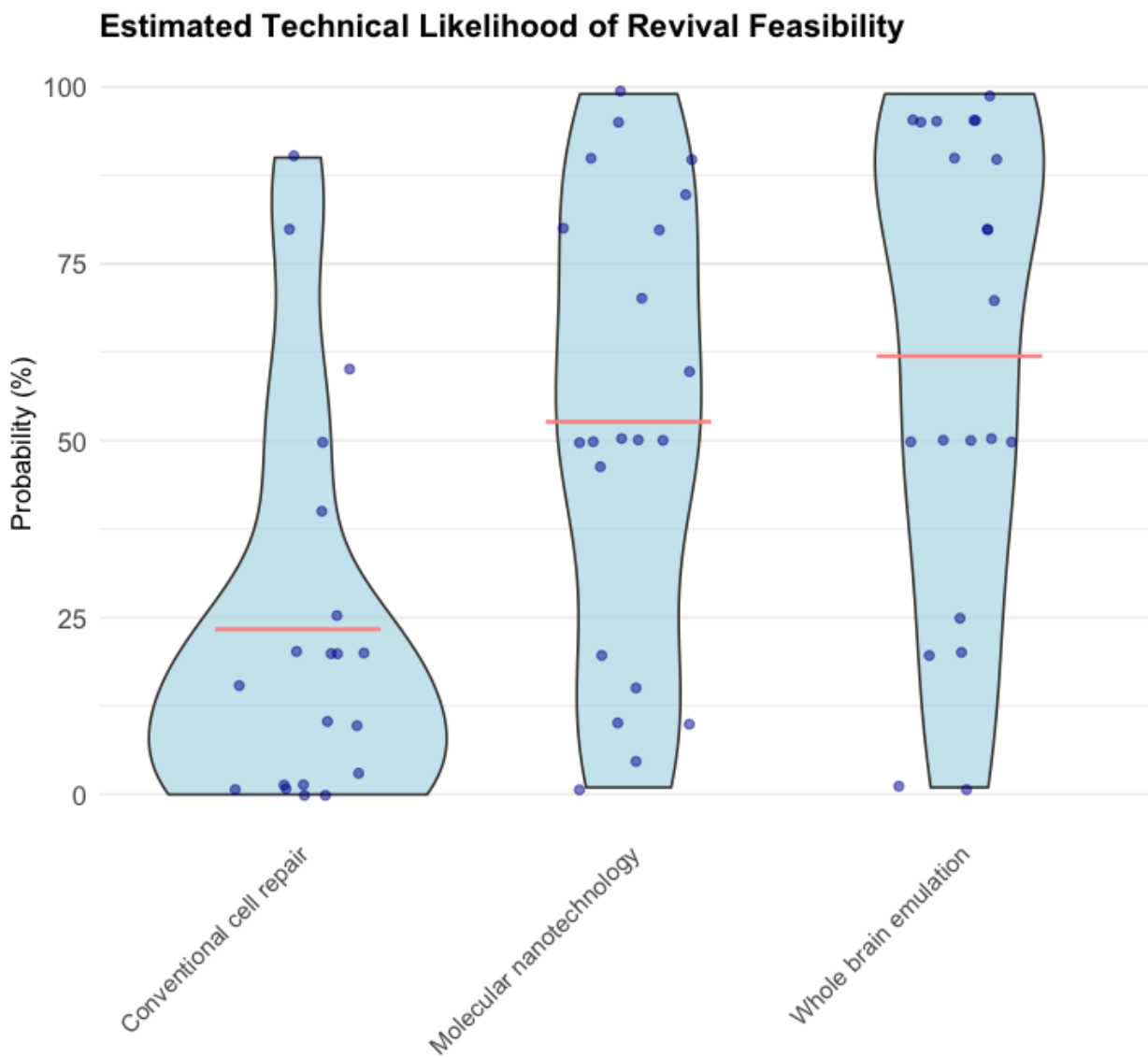

**Figure 13**. Estimated technical feasibility of different revival methods for people preserved in 2025 and earlier. Violin plots showing the distribution of probability estimates (0-100%) for three proposed revival technologies: conventional cell repair, molecular nanotechnology, and whole brain emulation. Individual estimates shown as horizontally jittered blue dots with horizontal red lines indicating means.

We also asked respondents the following:

> If you think that multiple of the proposed revival methods might one day be possible, which method do you think would be available first?

We note that this was a differently framed question than the previous question, because it did not specify when the preservation was performed. In the future, it could be split into two questions – one about which method would be available first (if any) for people preserved up until the present day, and one about which would be available for future hypothetical long-term preservation procedures (again, if any). Among the three options, 5 (23%) thought that conventional cell repair would be available first, 2 (9.1%) expected that molecular nanotechnology would, 14 (64%) selected whole brain emulation (WBE), 1 (4.5%) wrote "WBE at least for people preserved before 2025," and 1 (4.5%) was unsure.

Next, we discuss each potential revival method in more depth and discuss why respondents may have made these selections.

*Conventional cell repair*

Numerous approaches have been proposed over the years for the *in situ* repair of tissues to reverse damage due to the preservation process and revive the person. The first formal proposal for cellular repair of biostasis patients we are aware of is Jerome White's 1969 conference paper, "Viral Induced Repair of Damaged Neurons with Preservation of Long Term Information Content" (de Wolf, 2017; White, 1969). White's proposal was to use artificially constructed virus particles to introduce genetic repair programs into damaged cells. To preserve long-term memory content during cellular repair, he suggested that repair programs could "incorporate appropriate RNA tapes into itself upon entry and release them on termination of repair" to maintain information stored in hypothetical molecular feedback cycles involving mRNA. While this was a detailed, information-centric proposal that was notable for its era, most views in neuroscience these days favor structural connectivity patterns over molecular feedback cycles as the primary basis for long-term memory storage. Moreover, it is unclear how this approach would be sufficient to address the many forms of molecular and cellular damage that occur during preservation and associated ischemia.

Since White's proposal, several other similar proposals for *in situ* repair, with different levels of detail, have also been described (Darwin, 1977; Drexler, 1981; Donaldson, 1988; Darwin, 1988; Wowk, 1988; Merkle, 1994). In recent years, Robert Freitas has written a detailed proposal for *in situ* repair as a revival method for biostasis patients, which he calls "conventional cell repair" (R. Freitas, 2022). As a brief summary, his approach proposes to excavate material from blood vessels and insert a proposed device called a "vasculoid," which would be a mechanical, cilia-like system that allows the transport of material in the vasculature system. Using this vasculoid, sensors would scan the tissue from within the vasculature, to a resolution of ~100 nm feature resolution in three dimensions. Based on this scanned data, a repair plan would be computed. Nanorobots would then be deployed to physically carry out the task of replacing damaged and extracted molecules. It is important to note that this remains a theoretical concept, as the required nanotechnology does not yet exist and would require significant technological advances beyond current capabilities.

A key constraint that Freitas acknowledges with conventional cell repair is that the key structural variability in the brain that encodes personal identity, including long-term memories, must be at the level of 100 nm or higher (R. Freitas, 2022). This would include structural features such as the basic structure of dendritic spines and synapses. However, it is also possible that smaller substructural features, such as individual molecules or molecular complexes of the synaptic vesicle pools, active zones, or postsynaptic densities, cannot be inferred and must be imaged directly in order to capture key variability necessary for personal identity. In that case, the sensors would not be able to map the necessary information at a sufficient resolution with this method. For this reason, as well as the relative lack of precedent for this type of technology, respondents may have been more skeptical of this method being technically feasible.

*Molecular nanotechnology-based reconstruction*

An alternative approach that has been proposed for revival is to disassemble the preserved person's body – and in particular their brain – via a layer-by-layer scan of the atoms or molecules (Merkle, 1994; R. Freitas, 2022). After disassembly, there are two ways that the atoms or molecules could be restored. First, they could be replaced by new versions, for example by "3d printing" a new body with the same pattern. Second, they could be stored in a "filing cabinet," chemically altered if necessary for repairs, and then moved back into their original positions relative to one another. The method of keeping the original atoms is expected to be slower and more expensive (R. Freitas, 2022).

Important advantages of this molecular nanotechnology-based reconstruction approach compared to approaches that rely on scanning tissue through the vasculature are that (a) it would allow for much more detailed scans – by approximately 1000-fold – and (b) it would allow for the sensing and manipulation of individual atoms and chemical moieties during repair.

While a brute force approach such as molecular nanotechnology-based reconstruction seems to not break known laws of physics, and could theoretically be accomplished through the use of technologies evolving from today's atomic force microscopy, it would clearly require significant technological advances well, well beyond our current capabilities. Some may have been skeptical of the technical feasibility for this reason. However, in a world of accelerating technological advances, it is very difficult to predict what will or will not be possible in a century or two, assuming that humanity has not succumbed to civilizational collapse or extinction.

We also asked respondents:

> Do you think that molecular nanotechnology, if it is ever developed to the level suggested by Eric Drexler, Ralph Merkle, or Robert Freitas, would be compatible with the following preservation methods as they are available today?

The answer choices were "Very likely", "Likely", "Unsure", "Unlikely", and "Very unlikely." Practitioners expressed optimism about its potential compatibility with both pure cryopreservation and ASC, with 8 (38%) rating it as "Very likely" compatible with pure cryopreservation and 12 (57%) giving the same assessment for ASC (**Figure 14**). Differences between the methods were small and may be noise, but it is also possible that some respondents gave higher probability estimates for ASC because they expected that it has better structural information retention than pure cryopreservation, at least in their contemporary forms. However, this is confusing because respondents may have interpreted the word "compatible" differently – some as asking about structural preservation quality, others about whether MNT could physically work with tissue preserved by each method. In future versions of the questionnaire, this question about compatibility should be clarified to specifically ask whether MNT could physically repair or reconstruct tissue preserved by each method, instead of being conditional on their relative information preservation. Regardless, when combining "Very likely" and "Likely" responses, a clear majority of respondents viewed both preservation methods as likely compatible with MNT.

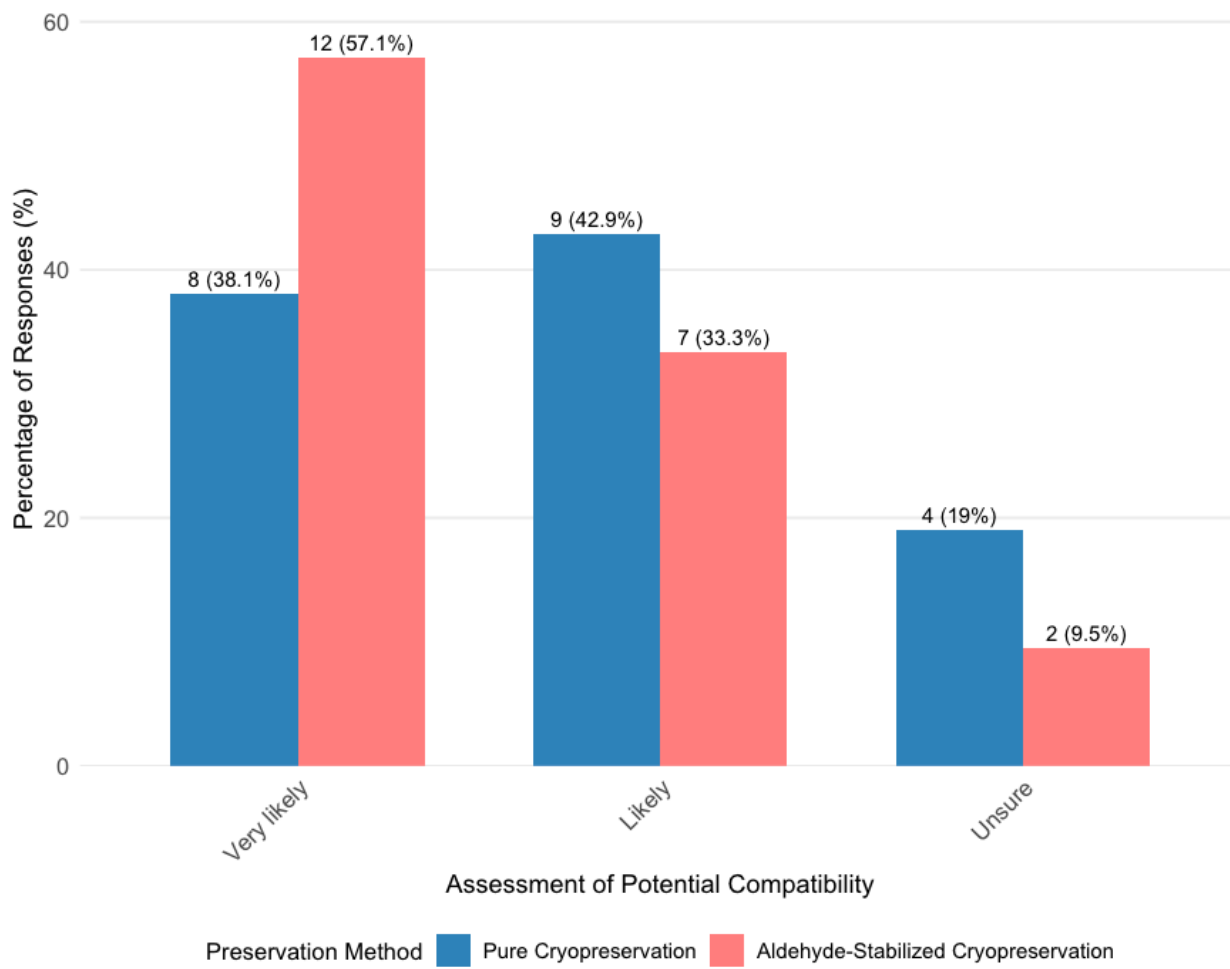

**Figure 14**. Theoretical compatibility of molecular nanotechnology (MNT) with different preservation methods. Bar plot showing the percentage of respondents rating the compatibility as "Very likely," "Likely," or "Unsure" for pure cryopreservation (blue) and aldehyde-stabilized cryopreservation (red). No respondents chose "Unlikely" or "Very unlikely" for either option.

Some in the biostasis community have expressed concerns that crosslinking of molecules would not be compatible with biological revival. However, molecular nanotechnology-based reconstruction presupposes a powerful, "brute force" technology that would most likely be sufficient to reverse any such molecular damage, as long as the information were still intact (McKenzie et al., 2024c). Our respondents' opinions are consistent with the stated views of Eric Drexler (Drexler, 1986), Ralph Merkle (Merkle, 1994), and Robert Freitas (R. A. Freitas, 2022, p. 482), each of whom have written or implied that MNT-based reconstruction could theoretically be compatible with the cross-linking fixatives used in aldehyde preservation.

We also asked respondents about their self-assessed knowledge levels of MNT. Our respondents gave high likelihoods of both preservation methods being most likely compatible with MNT – that is, if MNT is ever developed – across all self-assessed knowledge levels of MNT. For example, among the five respondents who self assessed their knowledge of MNT as "High", all five estimated it was "Very likely" to be compatible with both pure cryopreservation and ASC, and four out of five estimated it was "Very likely" to be compatible with pure cryopreservation, with the other expecting that it was "Likely" to be compatible with pure cryopreservation.

*Whole brain emulation*

Proposals for whole brain emulation (WBE) as a revival method generally involve mapping the detailed neural architecture of the preserved person's brain through scanning techniques, and then constructing a computational representation of the person. Once the model for this emulated person is created, the person could then be revived, for example either via connection to a biological or robotic body in the physical world or via a virtual body in a virtual world. One influential proposal for WBE following preservation was written by Kenneth Hayworth, who suggested a detailed method for embedding a fixed brain, slicing the brain, scanning the sections with electron microscopy, and emulating the brain *in silico* on the basis of this data (Hayworth, 2010). Notably, some have critiqued Hayworth's proposal because this type of scan would not have any molecular information.

There are many disagreements about the level of scanning detail required to preserve personal identity in a WBE procedure. This likely will affect how quickly WBE is developed. However, it does not seem likely to affect whether it will ever be developed. There are already technologies, such as immuno-electron microscopy or expansion microscopy, that could foreseeably serve as the precursors of more advanced technologies to map the locations of key (or even all) biomolecules directly.

Because it does seem not possible to definitively resolve the debates about what level of scanning detail will be needed for a future WBE procedure, most people believe that current biostasis preservation methods should be designed conservatively to maintain as much biomolecular structure and organization as possible, even if we cannot yet determine which specific molecular details will be essential for successful revival.

Many people have philosophical concerns about WBE, especially regarding whether such an emulation would allow personal psychological experience – which is often referred to as personal identity – to genuinely continue. This is what the vast majority of people pursuing biostasis desire. As a result, WBE is a highly polarizing topic within the field of biostasis. One survey of cryonicists found that approximately half of individuals felt that an emulated version of them would be them in the ways that matter (Marty, 2023). This is related to the teletransporter problem in the philosophical literature on personal identity (Wiley, 2023). A survey of philosophers on the teletransporter survival question found that 40.1% consider it death, 35.2% consider it survival, 10.1% are agonistic/undecided, 7.5% feel that there is no fact of the matter, and the rest have other views (Bourget and Chalmers, 2023).

We asked our respondents the following question:

> What do you think about the idea of whole brain emulation as a revival option for yourself? Imagine that it was done optimally – i.e. a full whole brain emulation, possibly using a type of neuromorphic computing if necessary – not just a partial brain simulation. Alternatively, if you don't think you yourself would ever be interested in biostasis, please think of one specific person you know who is signed up for biostasis and imagine how they might answer.

Of our respondents, 13 (59%) answered that they "Mostly or fully expect that it would be revival and allow for continuity of experience", 3 (14%) "Lean towards that it would be revival and allow for continuity of experience", 1 (5%) "Unsure", 1 (5%) "Lean towards that it would NOT be revival or allow for continuity of experience", and 4 (18%) "Mostly or fully expect that it would NOT be revival or allow for continuity of experience." The majority (73%) at least leaned towards considering WBE to be genuine revival, whereas 23% at least leaned towards not considering WBE to be genuine revival (**Figure 15**).

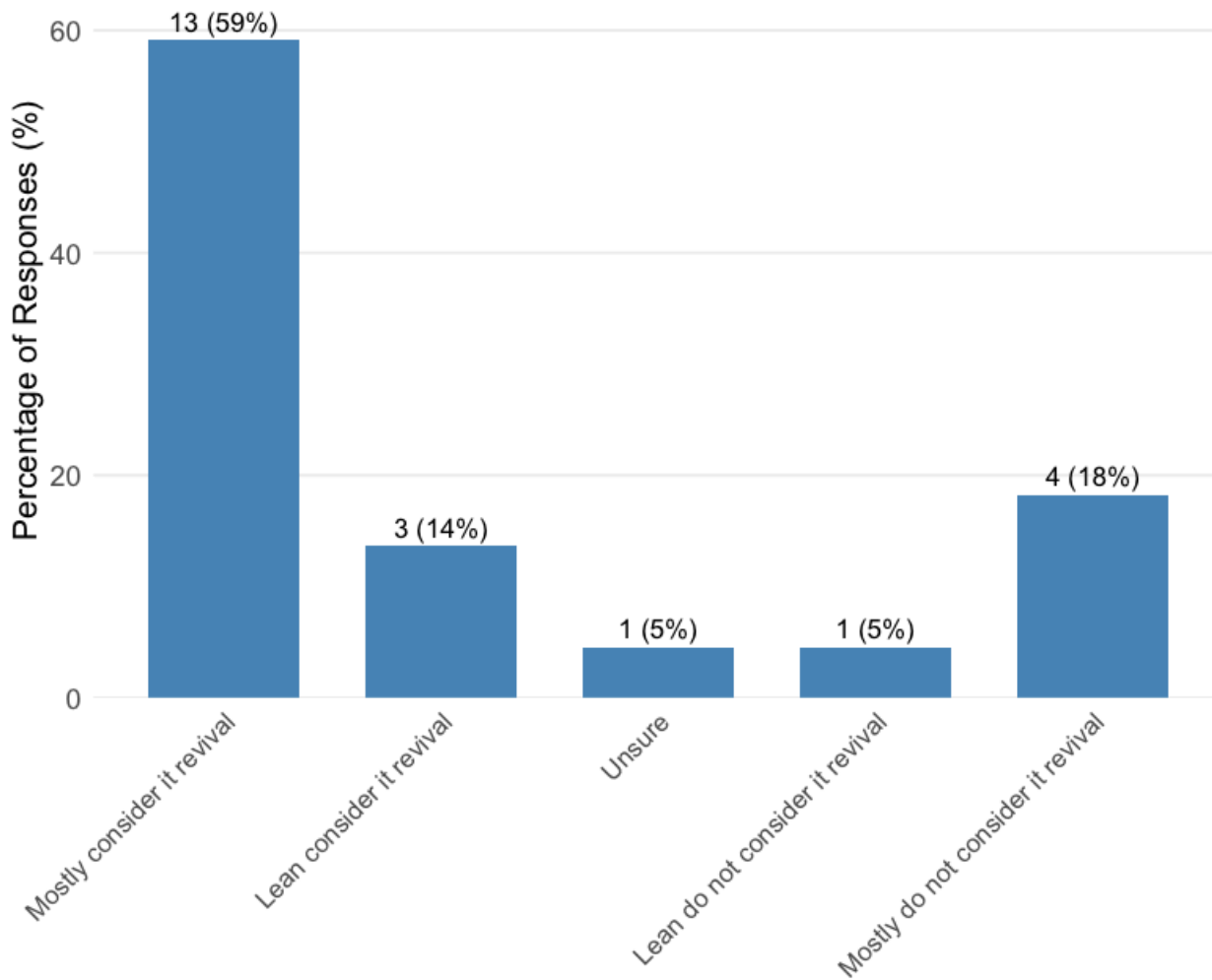

**Figure 15**. Practitioner views on whether whole brain emulation would constitute genuine revival. Bar plot showing the percentage of respondents across five categories ranging from those who mostly/fully expect that it would be revival and allow for continuity of experience to those who mostly/fully expect it would not.

We note several qualifications. First, "mostly" is not the same as "fully." Given the preponderance of answers in the "mostly or fully" category, these should potentially be split up in a subsequent version of the questionnaire. Second, "could" means that respondents might have specific requirements. Clearly, the fidelity of the emulation matters. The computational substrate could also matter. For example, some expect that the substrate might need to be neuromorphic (McKenzie et al., 2024b). Additionally, some proponents of WBE believe that any branch immediately becomes a distinct person whose termination would constitute death (i.e. branching identity) (Cerullo, 2015; Wiley, 2014), while others argue that personal identity consists primarily in memories, making the creation and deletion of branches potentially permissible in some contexts as long as memories persist (Hayworth, 2010).

One respondent noted that "Identity is as identity does," adding "A faithful emulation of an individual who was enthusiastic about survival by whole brain emulation would continue to express enthusiasm about it by definition of faithful emulation. Conversely, someone with philosophical reservations about it would be expected to express discomfort and unhappiness upon learning they were 'running' on an alternative substrate unless they had made prior peace with that as a possible outcome."

The fact that some consider WBE to be a revival strategy and others do not has the potential to cause confusion. For example, we asked one question about the degree of body repair or transplant that would be needed as a part of revival, but then many individuals were understandably confused about how this question would apply to WBE. Therefore, this question did not yield useful answers for analysis.

Notably, beyond metaphysical concerns, there are numerous other concerns with whole brain emulation. These include both broader socio-political concerns as well as personal safety. Just because some respondents answered that they thought WBE would constitute revival does not mean necessarily that they think it would be a desirable outcome for themselves or a good idea for society, nor that we are necessarily in a position to evaluate this today. In fact, personal safety concerns for WBE – such as having your mind run on an operating system controlled by someone not necessarily having your best interests in mind (Thau, 2020) – might be more concerning if one does hold the view that an emulated version of them would "be them."

**Suggestions for improvements in the field**

*Increasing public interest and legitimacy*

Multiple respondents described the need to increase the scientific credibility and medical legitimacy of biostasis. Respondents called for "rigorous publication of data studies," "frequent peer review," and demonstration of "scientific rigor" to distinguish the field from perceptions of "snake oil." Several respondents suggested positioning biostasis within broader medical contexts, such as organ banking and transplantation medicine. One noted that "organ cryopreservation for transplants was first looked upon with skepticism and ethical concerns but are now widely accepted." Others recommended building upon potentially increasing interest in rejuvenation biotechnology in the future, arguing that "widespread belief in the possibility of indefinite lifespans" would make biostasis more appealing by making extended life seem more plausible.

*Research collaboration and infrastructure*

Several respondents identified barriers to collaboration, noting that "the number of labs is very small, and they compete rather than collaborate." Proposed solutions to this problem included establishing "common standards" and "standardized protocols," creating a shared database for communicating current projects, and organizing regular conferences, such as online annual meetings, to facilitate interaction between geographically distant research groups. One respondent suggested a "joint donor network among labs" to maximize the usefulness of donated anatomical tissue across different groups with varying research focuses. As one practical response to these suggestions, we plan on starting a shared, publicly accessible research project database on our Biostasis Roadmap website (https://www.biostasis.xyz), where people interested in the field can potentially find others to collaborate with or learn about their progress. This will hopefully allow more individuals to work together on the forefront of research projects.

*Funding challenges and strategies*

Respondents who answered the optional question about biostasis research funding universally agreed that funding is "insufficient" or that the field is "tragically under-funded," with several noting that ideally there would be NIH-level governmental support. In fairness, it is not particularly surprising that the biostasis research community feels that it is receiving an insufficient level of funding, as this is probably true of the vast majority of research communities with respect to their own fields of inquiry. Strategic recommendations to increase funding included framing research to highlight its relevance to "broader medical or scientific applications" to access larger funding pools, targeting wealthy older individuals through "really effective fundraisers," and pursuing "moonshot projects" or government-sponsored initiatives. Multiple respondents suggested that achieving reversible preservation milestones would be very useful for attracting serious funding attention from traditional biomedical research sources.

**Limitations of this questionnaire**

This manuscript has several limitations that should be considered when interpreting the data. First, the findings are clearly subject to selection bias. The n = 22 respondents were recruited through the speakers at a biostasis conference and their associated professional networks. This creates the potential for over-representing certain types of biostasis researchers. And of course, biostasis practitioners themselves are likely to be relatively optimistic about prospects in the field, both due to financial incentives and as a selection effect of the fact that they chose to join the field. Additionally, the careers of some of the researchers involved may depend upon continued research funding, presumably making them less likely to say "no more research is needed."

Second, another limitation is the potential for social desirability bias. That is, respondents may have been influenced by their professional relationships or desire to provide answers they believed would be viewed favorably by the research team or the broader biostasis community. We attempted to mitigate this by emphasizing that participants should simply provide their best guesses. Moreover, those interested in biostasis are not known for being especially agreeable. Still, social desirability bias has a clear potential influence on the results and should be kept in mind.

Third, the questionnaire format often left it unclear why respondents provided certain answers. While we collected probability estimates and categorical responses, we did not systematically probe the underlying reasoning, assumptions, or evidence base that informed these judgments. Future work could benefit from more structured follow-up questions or interviews to better understand the rationale behind practitioner estimates.

Finally, the uncertainty in predicting technological timelines decades or centuries into the future makes these estimates, especially the timeline ones, highly speculative. Respondents must simultaneously consider the technical difficulty of the problems, their estimates of the rate of scientific progress in general (including AI, which can vary widely), the level of investment they expect to be put into particular sub-aspects of biostasis, as well as societal factors such as whether the stigma towards biostasis diminishes over time. These factors could make predictions highly variable across practitioners even if they may agree on many of the technical aspects of the question. This uncertainty should be kept in mind when interpreting the responses.

**Conclusions**

Our manuscript provides an examination of current practitioner perspectives in biostasis. We note several key findings. First, most respondents believe that provably reversible mammalian cryopreservation is most likely decades away, and that this milestone will take an even longer time to be demonstrated in humans. This finding may help to temper potentially unrealistic expectations among some lay people on this topic. Second, respondents see ideal-case preservation quality, geographic barriers, and procedural execution as the most likely failure modes in contemporary practice. Third, there are several technical disagreements among respondents, such as the optimal perfusion routes and the adequacy of immersion fixation for preserving synaptic connectivity across the brain. These questions may be amenable to empirical testing. Fourth, respondents believe that whole brain emulation is the most likely revival method to be developed first, as well as the most technically feasible method on people preserved today. However, respondents are divided on the metaphysical question of whether whole brain emulation would constitute genuine revival. Fifth, molecular nanotechnology is also seen as relatively likely to be technically feasible, and compatible with both pure cryopreservation and aldehyde-based preservation methods. On the other hand, the proposal for conventional cell repair as detailed by Robert Freitas is viewed more skeptically, at least in terms of its applicability for people preserved in 2025 and earlier. Taken together, these findings identify areas for future empirical research and provide baseline data for tracking the field's evolution. Subsequent questionnaires could potentially expand on these initial results to monitor any changes in perspectives as the field develops.

**Abbreviations**

**ASC**: Aldehyde-Stabilized Cryopreservation, **BBB**: Blood-Brain Barrier, **BPF**: Brain Preservation Foundation, **EEG**: Electroencephalography, **FDA**: Food and Drug Administration, **ITS**: Intermediate Temperature Storage, **MNT**: Molecular Nanotechnology, **NIH**: National Institutes of Health, **WBE**: Whole Brain Emulation.

**Acknowledgments**

We thank all respondents for their valuable time and insights. We also thank the organizers of the Biostasis Week Conference 2025 at Vitalist Bay for facilitating participant recruitment.We thank all respondents for their valuable time and insights. We also thank the organizers of the Biostasis Week Conference 2025 at Vitalist Bay for facilitating participant recruitment.

**Data availability**

Anonymized data and analysis code are available here: https://github.com/andymckenzie/biostasis_forecasts.

**Conflict of interest**

**References**

21CM Cryopreservation Eval Page – The Brain Preservation Foundation [WWW Document], 2015. . Brain Preservation Foundation. URL https://www.brainpreservation.org/21cm-cryopreservation-eval-page/ (accessed 6.3.25).
Ames, A., Wright, R.L., Kowada, M., Thurston, J.M., Majno, G., 1968. Cerebral ischemia. II. The no-reflow phenomenon. Am J Pathol 52, 437–453.
Best, B.P., 2008. Scientific justification of cryonics practice. Rejuvenation Res 11, 493–503. https://doi.org/10.1089/rej.2008.0661
Biostasis Week Conference (May 17-18) @ Vitalist Bay [WWW Document], 2025. URL https://www.vitalistbay.com/biostasis (accessed 5.3.25).
Bourget, D., Chalmers, D.J., 2023. Philosophers on Philosophy: The 2020 Philpapers Survey. Philosophers' Imprint 23. https://doi.org/10.3998/phimp.2109
Cerullo, M.A., 2015. Uploading and Branching Identity. Minds & Machines 25, 17–36. https://doi.org/10.1007/s11023-014-9352-8
Darwin, M., 1988. Resuscitation: A Speculative Scenario for Recovery. Cryonics Magazine 33–37.
Darwin, M.G., 1977. The anabolocyte: a biological approach to repairing cryoinjury. Life Extension Magazine 80–83.

de Wolf, A., 2017. Forever Lost? The First Cryonics Brain Repair Paper [WWW Document]. Biostasis. URL https://www.biostasis.com/forever-lost-the-first-cryonics-brain-repair-paper/ (accessed 6.10.25).

de Wolf, A., Phaedra, C., Perry, R.M., Maire, M., 2020. Ultrastructural Characterization of Prolonged Normothermic and Cold Cerebral Ischemia in the Adult Rat. Rejuvenation Research 23, 193–206. https://doi.org/10.1089/rej.2019.2225

Donaldson, T., 1988. 24th Century Medicine. Cryonics Magazine 9, 16–34.

Drexler, K.E., 1986. Engines of Creation. Anchor Press/Doubleday.

Drexler, K.E., 1981. Molecular engineering: An approach to the development of general capabilities for molecular manipulation. Proc Natl Acad Sci U S A 78, 5275–5278. https://doi.org/10.1073/pnas.78.9.5275

Freitas, R., 2022. Cryostasis Revival: The Recovery of Cryonics Patients through Nanomedicine. Cryonics Magazine 2nd Quarter 2022, 15–23.

Freitas, R.A., 2022. Cryostasis Revival: The Recovery of Cryonics Patients through Nanomedicine, First Edition. ed. Alcor Life Extension Foundation.

Garrood, M., Thorn, E.L., Goldstein, A., Sowa, A., Janssen, W., Wilson, A., López, C.S., Shankar, R., Stempinski, E.S., Farrell, K., Crary, J.F., McKenzie, A.T., 2025. Preservation of cellular structure via immersion fixation in brain banking. Free Neuropathol 6, 4. https://doi.org/10.17879/freeneuropathology-2025-6104

German, A., Akdaş, E.Y., Flügel-Koch, C., Fejtova, A., Winkler, J., Alzheimer, C., Zheng, F., 2025. Functional recovery of adult brain tissue arrested in time during cryopreservation by vitrification. bioRxiv: The Preprint Server for Biology 2025.01.22.634384. https://doi.org/10.1101/2025.01.22.634384

Harris, S., 1989. Will Cryonics Work? Examining the Probabilities. Cryonics Magazine 10, 36–48.

Hayworth, K., 2018. Letter of support for Aldehyde Stabilized Cryopreservation to be developed into a medical procedure [WWW Document]. Brain Preservation Foundation. URL https://www.brainpreservation.org/wp-content/uploads/2018/02/letterofsupportforasc_kennethhayworth_jan2018-signed.pdf (accessed 6.3.25).

Hayworth, K., 2011. The Brain Preservation Technology Prize: A challenge to cryonicists, a challenge to scientists. Cryonics Magazine 32, 5–9.

Hayworth, K., 2010. Killed by Bad Philosophy [WWW Document]. Brain Preservation Foundation. URL https://www.brainpreservation.org/content-2/killed-bad-philosophy/ (accessed 4.9.24).

Januszewski, M., Templier, T., Hayworth, K., Peale, D., Hess, H., 2025. Accelerating Neuron Reconstruction with PATHFINDER. bioRxiv 2025.05.16.654254. https://doi.org/10.1101/2025.05.16.654254

Kloner, R.A., King, K.S., Harrington, M.G., 2018. No-reflow phenomenon in the heart and brain. Am J Physiol Heart Circ Physiol 315, H550–H562. https://doi.org/10.1152/ajpheart.00183.2018

Krassner, M.M., Kauffman, J., Sowa, A., Cialowicz, K., Walsh, S., Farrell, K., Crary, J.F., McKenzie, A.T., 2023. Postmortem changes in brain cell structure: a review. Free Neuropathol 4, 4–10. https://doi.org/10.17879/freeneuropathology-2023-4790

Marty, M., 2023. The Cryonics Survey of 2022, part 2. Frozen Futures. URL https://frozenfutures.substack.com/p/the-cryonics-survey-of-2022-part-5ed (accessed 4.9.24).

Mazur, A., Ayyadevara, S., Mainali, N., Patchett, S., Uden, M., Roa, R.I., Fahy, G.M., Shmookler Reis, R.J., 2024. Model biological systems demonstrate the inducibility of pathways that strongly reduce cryoprotectant toxicity. Cryobiology 115, 104881. https://doi.org/10.1016/j.cryobiol.2024.104881

McFadden, W.C., Walsh, H., Richter, F., Soudant, C., Bryce, C.H., Hof, P.R., Fowkes, M., Crary, J.F., McKenzie, A.T., 2019. Perfusion fixation in brain banking: a systematic review. Acta Neuropathol Commun 7, 146. https://doi.org/10.1186/s40478-019-0799-y

McIntyre, R.L., Fahy, G.M., 2015. Aldehyde-stabilized cryopreservation. Cryobiology 71, 448–458. https://doi.org/10.1016/j.cryobiol.2015.09.003

McKenzie, A.T., Keberle, A., Minerva, F., Zeleznikow-Johnston, A., Harrow, J., 2025. The Right to Immediate Preservation: Addressing Legal Barriers Due to Death Investigation. Forensic Sciences 5, 16. https://doi.org/10.3390/forensicsci5020016

McKenzie, A.T., Nnadi, O., Slagell, K.D., Thorn, E.L., Farrell, K., Crary, J.F., 2024a. Fluid preservation in brain banking: a review. Free Neuropathol 5, 5–10. https://doi.org/10.17879/freeneuropathology-2024-5373

McKenzie, A.T., Wowk, B., Arkhipov, A., Wróbel, B., Cheng, N., Kendziorra, E.F., 2024b. Biostasis: A Roadmap for Research in Preservation and Potential Revival of Humans. Brain Sci 14, 942. https://doi.org/10.3390/brainsci14090942

McKenzie, A.T., Zeleznikow-Johnston, A., Sparks, J.S., Nnadi, O., Smart, J., Wiley, K., Cerullo, M.A., de Wolf,

A., Minerva, F., Risco, R., Church, G.M., de Magalhães, J.P., Kendziorra, E.F., 2024c. Structural brain preservation: a potential bridge to future medical technologies. Front. Med. Technol. 6. https://doi.org/10.3389/fmedt.2024.1400615

Merkle, R., 1994. The Molecular Repair of the Brain. Cryonics Magazine 15, 16–31.

Merkle, R.C., 1992. The technical feasibility of cryonics. Med Hypotheses 39, 6–16. https://doi.org/10.1016/0306-9877(92)90133-w

More, M., 2023. What is the probability of cryonics working? [WWW Document]. The Biostasis Standard. URL https://biostasis.substack.com/p/what-is-the-probability-of-cryonics (accessed 7.1.25).

More, M., 1995. The Diachronic Self (PhD Thesis). University of Southern California.

Norton, B., 2020. Cryonics Calculator [WWW Document]. cryonicscalculator. URL https://www.cryonicscalculator.com (accessed 5.30.25).

Osetrova, M., Tkachev, A., Mair, W., Guijarro Larraz, P., Efimova, O., Kurochkin, I., Stekolshchikova, E., Anikanov, N., Foo, J.C., Cazenave-Gassiot, A., Mitina, A., Ogurtsova, P., Guo, S., Potashnikova, D.M., Gulin, A.A., Vasin, A.A., Sarycheva, A., Vladimirov, G., Fedorova, M., Kostyukevich, Y., Nikolaev, E., Wenk, M.R., Khrameeva, E.E., Khaitovich, P., 2024. Lipidome atlas of the adult human brain. Nat Commun 15, 4455. https://doi.org/10.1038/s41467-024-48734-y

Park, J., Papoutsi, A., Ash, R.T., Marin, M.A., Poirazi, P., Smirnakis, S.M., 2019. Contribution of apical and basal dendrites to orientation encoding in mouse V1 L2/3 pyramidal neurons. Nat Commun 10, 5372. https://doi.org/10.1038/s41467-019-13029-0

Poo, M., Pignatelli, M., Ryan, T.J., Tonegawa, S., Bonhoeffer, T., Martin, K.C., Rudenko, A., Tsai, L.-H., Tsien, R.W., Fishell, G., Mullins, C., Gonçalves, J.T., Shtrahman, M., Johnston, S.T., Gage, F.H., Dan, Y., Long, J., Buzsáki, G., Stevens, C., 2016. What is memory? The present state of the engram. BMC Biology 14, 40. https://doi.org/10.1186/s12915-016-0261-6

Shcherbakova, L.N., Tel'pukhov, V.I., Trenin, S.O., Bashikov, I.A., Lapkina, T.I., 1986. Permeability of the blood-brain barrier for intravascular formaldehyde. Bull Exp Biol Med 102, 1553–1554. https://doi.org/10.1007/BF00854688

Thau, T., 2020. Cryonics for all? Bioethics 34, 638–644. https://doi.org/10.1111/bioe.12710

Vita-More, N., Barranco, D., 2015. Persistence of Long-Term Memory in Vitrified and Revived Caenorhabditis elegans. Rejuvenation Res 18, 458–463. https://doi.org/10.1089/rej.2014.1636

White, J.B., 1969. Viral-Induced Repair of Damaged Neurons with Preservation of Long-Term Information Content. Presented at the Second Annual Conference of the Cryonic Societies of America, University of Michigan at Ann Arbor.

Wiley, K., 2023. Nondestructive Mind Uploading and the Stream of Consciousness. SSRN. https://doi.org/10.2139/ssrn.4560723

Wiley, K., 2018. Implications of the BPF large mammal brain preservation prize. Brain Preservation Foundation. URL https://www.brainpreservation.org/implications-of-the-bpf-large-mammal-brain-preservation-prize/ (accessed 11.13.23).

Wiley, K., 2014. A Taxonomy and Metaphysics of Mind-Uploading. Humanity+ Press and Alautun Press.

Wowk, B., 2025. A Suspended Animation Research Hierarchy. The Biostasis Standard. URL https://biostasis.substack.com/p/a-suspended-animation-research-hierarchy?utm_medium=email&triggerShare=true (accessed 6.3.25).

Wowk, B., 2022. Matters of Life and Death: Reflections on the Philosophy and Biology of Human Cryopreservation. Cryonics Magazine 43, 3–14.

Wowk, B., 2011. Systems for Intermediate Temperature Storage for Fracture Reduction and Avoidance. Cryonics Magazine 32, 7–13.

Wowk, B., 1988. Cell Repair Technology. Cryonics Magazine 9, 21–30.